\documentclass{article}
\usepackage{amssymb,amsmath}
\usepackage{epsfig}
\usepackage{pdflscape}
\usepackage{subfigure}
\usepackage{color}
\usepackage{lscape}
\usepackage{bbm}
\usepackage{bm}
\usepackage{mathrsfs}
\usepackage{amsfonts}
\usepackage{graphics}
\usepackage{indentfirst}
\usepackage[square,numbers,sort&compress]{natbib}
\usepackage{multicol}

\newtheorem{rem}{Remark}[section]
\newcommand{\br}{\begin{rem}}
\newcommand{\er}{\end{rem}}
\newtheorem{ex}[rem]{Example}
\newcommand{\bex}{\begin{ex}}
\newcommand{\eex}{\end{ex}}
\newtheorem{Def}[rem]{Definition}
\newcommand{\bd}{\begin{Def}}
\newcommand{\ed}{\end{Def}}
\newtheorem{theorem}[rem]{Theorem}
\newcommand{\bt}{\begin{theorem}}
\newcommand{\et}{\end{theorem}}
\newtheorem{Prop}[rem]{Proposition}
\newcommand{\bp}{\begin{Prop}}
\newcommand{\ep}{\end{Prop}}
\newtheorem{lemma}[rem]{Lemma}
\newcommand{\bl}{\begin{lemma}}
\newcommand{\el}{\end{lemma}}
\newcommand{\be}{\begin{equation}}
\newcommand{\ee}{\end{equation}}
\newcommand{\bea}{\begin{eqnarray}}
\newcommand{\eea}{\end{eqnarray}}
\newcommand{\pa}{\partial}
\newcommand{\nn}{\nonumber}
\newcommand{\adots}{\mathinner{\mkern2mu\raise1pt\hbox{.}\mkern2mu
\raise4pt\hbox{.}\mkern2mu\raise7pt\hbox{.}\mkern1mu}}

\title{Superintegrable Systems on 3 Dimensional\\ Conformally Flat Spaces}

\author{Allan P. Fordy\thanks{School of Mathematics,
University of Leeds, Leeds LS2 9JT, UK. ~~E-mail: a.p.fordy@leeds.ac.uk}
$\,$ and Qing Huang\thanks{School of Mathematics, Northwest University, Xi’an 710069,
People’s Republic of China ~~E-mail: hqing@nwu.edu.cn}
}

\begin{document}

\maketitle

\begin{abstract}
We consider Hamiltonians associated with 3 dimensional conformally flat spaces, possessing 2, 3 and 4 dimensional isometry algebras.  We use the conformal algebra to build additional {\em quadratic} first integrals, thus constructing a large class of superintegrable systems and the complete Poisson algebra of first integrals.  We then use the isometries to reduce our systems to 2 degrees of freedom.  For each isometry algebra we give a {\em universal} reduction of the corresponding general Hamiltonian.  The superintegrable specialisations reduce, in this way, to systems of Darboux-Koenigs type, whose integrals are reductions of those of the 3 dimensional system.
\end{abstract}



\section{Introduction}

On spaces of constant curvature, the geodesic equations automatically have higher order integrals, which are just built out of first order integrals, corresponding to the abundance of Killing vectors.  The construction of these higher order integrals is the first step towards constructing first integrals for Hamiltonians in ``natural form'' (the sum of kinetic and potential energies), with the kinetic energy being {\em quadratic} in momenta.
The simplest case is that involving {\em quadratic} integrals, which also happen to be important in the discussion of separable systems.  Thus for Hamiltonians associated with spaces of constant curvature, the task of building quadratic or higher order integrals is relatively simple.

This is no longer true for general conformally flat spaces, but in this case there is a large algebra of {\em conformal} symmetries.  We can then use the method introduced in \cite{f19-3} to build homogeneously quadratic integrals from {\em conformal symmetries}.  This is for the geodesic case (purely {\em kinetic} energy) and again the {\em first step} towards building integrals for Hamiltonians with potentials.  We are particularly interested in building enough first integrals, with appropriate Poisson relations, for complete integrability or superintegrability (see \cite{13-2} for a general review of superintegrability).  In \cite{f19-3}, the method was illustrated by reproducing some well known, {\em non-constant curvature} systems in 2 degrees of freedom, such as the Darboux-Koenigs systems \cite{72-5,02-6,03-11}, with one linear and two quadratic integrals, and a case from the classification \cite{11-3} of systems with one linear and a {\em cubic} integral.  In \cite{f19-2}, we applied this method to {\em non-constant curvature} systems in 3 degrees of freedom,  seeking systems with one {\em linear} and three {\em quadratic} integrals (since we were interested in maximally superintegrable systems).  We found several 3 parameter systems, corresponding to inequivalent choices of Killing vector (first order integral), but, without further restriction, the full Poisson algebra is very difficult to determine.

In \cite{f19-2}, we found that, by restricting parameters, we could increase the dimension of the isometry algebra of our metric, which enabled us to derive the {\em full} Poisson algebra of first integrals. In the present paper we also consider the case of 3 degrees of freedom, but now {\em assume} the existence of a larger isometry algebra, whilst not having ``too many'' isometries, since we wish to avoid constant curvature.  In 2 dimensions, Darboux and Koenigs {\em proved} that we can only have 1 isometry (ie 2 implies 3). In 3 dimensions we can have up to 4 isometries (since 5 implies 6). This ``gap phenomenon'' occurs in all dimensions and is the subject of \cite{14-7}.  In the current paper, we consider 2, 3 and 4 dimensional isometry algebras and derive the corresponding {\em general} Hamiltonian systems, with restrictions possessing further quadratic integrals.

In Section \ref{conformal} we give our basis of the conformal algebra and describe some involutive automorphisms, which play a very important role in our calculations.  We enumerate all subalgebras of dimension 2,3 and 4, and then the restrict to a smaller list, after taking equivalence with respect to the involutions.  We then give the general Hamiltonian (of type (\ref{3d-gen})) which is invariant with respect to each of these algebras.

In Section \ref{quadrat-int}, we describe our main approach to finding first integrals (the method introduced in \cite{f19-3}).  The existence of larger (than 1D) isometry algebras makes it possible to make a systematic derivation.  The existence of isometries also enables us to reduce from 3 to 2 degrees of freedom.  Our general approach to this was introduced in \cite{f18-2} and described in Section \ref{sec:reductions}.

Our main results are in Sections \ref{2D-super}-\ref{4D-super-reduce}.  In Section \ref{2D-super} we present all superintegrable systems in our class, associated with the two inequivalent isometry algebras of dimension 2.  For each of the resulting seven cases, we present a 4 dimensional Poisson algebra of linear and quadratic integrals.

In Section \ref{3D-super} we consider the three inequivalent isometry algebras of dimensions 3 and present all superintegrable systems in our class.  There are six examples, but several of these already appeared in \cite{f19-2}, so we only give brief details, except where we have made significant simplifications or where the case is new.

In Sections \ref{2D-super-reduce} and \ref{3D-super-reduce}, we consider the reductions of the systems described in Sections \ref{2D-super} and \ref{3D-super}.  We give {\em universal} reduction schemes, which only depend on the specific isometry algebra.  We give the general form of reductions in each case, together with a commuting integral for the general case, thus rendering the 2 dimensional system {\em Liouville integrable}.  When we adapt these reductions to our superintegrable cases, they reduce to systems of Darboux-Koenigs type (whenever the additional integrals can be reduced).  Our labelling of Darboux-Koenigs systems is taken from \cite{02-6,03-11}.

In Section \ref{4D-super-reduce} we consider the two 4D isometry algebras that appear in our classification.  Having a 4D isometry algebra is very restrictive, so leads to two specific Hamiltonians.  These can be thought of as specialisations of some of the general classes with smaller isometry algebras.  These specific Hamiltonians reduce to flat and constant curvature systems.


\section{The 3D Euclidean Metric and its Conformal Algebra}\label{conformal}

Consider metrics which are conformally related to the standard Euclidean metric in 3 dimensions, with Cartesian coordinates $(q_1,q_2,q_3)$.  The corresponding kinetic energy takes the form
\be\label{3d-gen}
H = \varphi(q_1,q_2,q_3) \left(p_1^2+p_2^2+p_3^2\right).
\ee
A {\em conformal invariant} $X$, linear in momenta, will satisfy $\{X,H\}=\lambda(X) H$, for some function $\lambda(X)$.  The conformal invariants form a Poisson algebra, which we call the {\em conformal algebra}. For special cases of $\varphi(q_1,q_2,q_3)$ there will be a subalgebra for which $\{X,H\}=0$, thus forming {\em true invariants} of $H$.  These correspond to {\em infinitesimal isometries} (Killing vectors) of the metric.  Constant curvature metrics possess $\frac{1}{2} n(n+1) = 6$ Killing vectors (when $n=3$).

The conformal algebra of this 3D metric has dimension $\frac{1}{2} (n+1)(n+2)=10$ (when $n=3$).  A convenient basis is as follows
\begin{subequations}\label{g1234}
\bea
&&  e_1 = p_1,\quad h_1 = -2(q_1p_1+q_2p_2+q_3p_3),\quad f_1=(q_2^2+q_3^2-q_1^2)p_1-2q_1q_2p_2-2q_1q_3p_3,  \label{g1}\\
&&  e_2 = p_2,\quad h_2=2(q_1p_2-q_2p_1),\quad f_2 =-4q_1q_2p_1-2(q_2^2-q_1^2-q_3^2)p_2-4q_2q_3p_3,  \label{g2}\\
&&  e_3 = p_3,\quad h_3 = 2(q_1p_3-q_3p_1),\quad f_3=-4q_1q_3p_1-4q_2q_3p_2-2(q_3^2-q_1^2-q_2^2)p_3,  \label{g3}\\
&&  h_4 = 4q_3p_2-4q_2p_3.   \label{g4}
\eea
\end{subequations}
The Poisson relations of the ten elements in the conformal algebra (\ref{g1234}) are given in Table \ref{Tab:g1234}.
\begin{table}[h]\centering
\caption{The 10-dimensional conformal algebra (\ref{g1234})}\label{Tab:g1234}\vspace{3mm}
{\footnotesize
\renewcommand\arraystretch{1.26}\begin{tabular}{|c||c|c|c||c|c|c||c|c|c||c|}
\hline   &$e_1$    &$h_1$     &$f_1$   &$e_2$   &$h_2$    &$f_2$    &$e_3$    &$h_3$   &$f_3$    &$h_4$\\[.10cm]\hline\hline
$e_1$    &0        &$2e_1$    &$-h_1$  &0       &$-2e_2$    &$-2h_2$   &0     &$-2e_3$ &$-2h_3$  &0\\[1mm]\hline
$h_1$    &   &0        &$2f_1$  &$-2e_2$ &0        &$2f_2$   &$-2e_3$  &0       &$2f_3$   &0\\[1mm]\hline
$f_1$    &   &   &0       &$-h_2$  &$-f_2$    &0        &$-h_3$  &$-f_3$  &0        &0\\[1mm]\hline\hline
$e_2$    &   &   &   &0       &$2e_1$   &$-2h_1$   &0   & 0       &$h_4$  &$4e_3$\\[1mm]\hline
$h_2$    &   &   &  &   &0        &$-4f_1$         & 0  &$-h_4$   &0   &$4h_3$\\[1mm]\hline
$f_2$    &   &  &   &   &         &0               &$h_4$ &0      &0   &$4f_3$\\[1mm]\hline\hline
$e_3$    &   &   &  &   &   &   &0        &$2e_1$  &$-2h_1$            &$-4e_2$\\[1mm]\hline
$h_3$    &   &   &   &    &   & &    &0       &$-4f_1$                 &$-4h_2$\\[1mm]\hline
$f_3$    &    &  &  &  &   &  &  &  &0                                 &$-4f_2$\\[1mm]\hline\hline
$h_4$    & &  & &   &   &    &    &  &    &0\\[1mm]\hline
\end{tabular}
}\end{table}
Note that this is an example of the conformal algebra given in \cite{f18-1} (Table 3), corresponding to the case $a_2=a_3=2,\; a_4=0$.  The subalgebra $\mathfrak{g}_1$, with basis (\ref{g1}) is just a copy of $\mathfrak{sl}(2)$.  We then make the {\em vector space decomposition} of the full algebra $\mathfrak{g}$ into invariant subspaces under the action of $\mathfrak{g}_1$:
$$
\mathfrak{g} = \mathfrak{g}_1 + \mathfrak{g}_2 + \mathfrak{g}_3 + \mathfrak{g}_4.
$$
The basis elements for $\mathfrak{g}_i$ have the same subscript and are given in the rows of (\ref{g1234}).

\begin{table}[h]
\begin{center}
\caption{The involutions of the conformal algebra (\ref{g1234})}\label{Tab:Invg1234}\vspace{3mm}
{\footnotesize\begin{tabular}{|c||c|c|c||c|c|c||c|c|c||c|}
\hline
         &$e_1$    &$h_1$     &$f_1$   &$e_2$   &$h_2$    &$f_2$  &$e_3$   &$h_3$    &$f_3$   &$h_4$   \\[.10cm]\hline\hline
$\iota_{12}$ &$e_2$ &$h_1$ &$\frac{1}{2}f_2$ &$e_1$ &$-h_2$ &$2 f_1$ &$e_3$ &$-\frac{1}{2}h_4$ &$f_3$ &$-2h_3$  \\[1mm]\hline
$\iota_{13}$ &$e_3$ &$h_1$ &$\frac{1}{2}f_3$ &$e_2$    &$\frac{1}{2}h_4$ &$f_2$ &$e_1$     &$-h_3$ &$2f_1$ &$2h_2$ \\[1mm]\hline
$\iota_{23}$ &$e_1$ &$h_1$ &$f_1$ &$e_3$    &$h_3$ &$f_3$ &$e_2$      &$h_2$ &$f_2$ &$-h_4$ \\[1mm]\hline
$\iota_{ef}$      &$-f_1$ &$-h_1$ &$-e_1$  &$-\frac{1}{2}f_2$ &$h_2$ &$-2e_2$   &$-\frac{1}{2}f_3$ &$h_3$ &$-2e_3$  &$h_4$\\[1mm]\hline
\end{tabular}}
\end{center}
\end{table}

The algebra (\ref{g1234}) possesses a number of involutive automorphisms:
\begin{equation*}
\begin{split}
&  \iota_{12}:\ (q_1,q_2,q_3)\mapsto(q_2,q_1,q_3),\quad \iota_{13}:\ (q_1,q_2,q_3)\mapsto(q_3,q_2,q_1),\quad \iota_{23}:\ (q_1,q_2,q_3)\mapsto(q_1,q_3,q_2),\\
&  \iota_{ef}:\ (q_1,q_2,q_3)\mapsto \left(-\frac{q_1}{q_1^2+q_2^2+q_3^2},-\frac{q_2}{q_1^2+q_2^2+q_3^2},-\frac{q_3}{q_1^2+q_2^2+q_3^2}\right),
\end{split}
\end{equation*}
whose action is given in Table \ref{Tab:Invg1234}.

\subsection{Low Dimensional Subalgebras of the Conformal Algebra $\mathfrak{g}$}\label{sec:subalgebras}

We are interested in subalgebras of $\mathfrak{g}$, having dimensions 2, 3 and 4.  Specifically, we enumerate all such subalgebras, with bases of the form $\left< K_1,\dots ,K_m\right>,\, m=2,3,4$, where $K_i$ are chosen from the list (\ref{g1234}).

\begin{table}[h]
\begin{center}
\caption{Subalgebras of $\mathfrak{g}$}\label{Tab:SubAlg}\vspace{3mm}
\begin{tabular}{|l|l|l|}
\hline
  Dimension    & Algebras             &Representative    \\[.10cm]\hline
2D  &$\langle e_i,h_1\rangle$, $\langle f_i,h_1\rangle$ for $i=1,2,3$  &  $\langle e_1,h_1\rangle$\\
       &$\langle e_i,e_j\rangle$, $\langle f_i,f_j\rangle$ for $1\leq i<j\leq3$  &  $\langle e_1,e_2\rangle$ \\
           & $\langle e_i,h_{5-i}\rangle$, $\langle f_i,h_{5-i}\rangle$ for $i=1,2,3$  &  $\langle e_1,h_{4}\rangle$  \\
      & $\langle h_i,h_1\rangle$ for $i=2,3,4$   &  $\langle h_1,h_2\rangle$  \\ \hline
3D   & $\langle e_1,e_2,e_3\rangle$, $\langle f_1,f_2,f_3\rangle$ &  $\langle e_1,e_2,e_3\rangle$ \\
  & $\langle e_i,h_1,f_i\rangle$ for $i=1,2,3$  & $\langle e_1,h_1,f_1\rangle$ \\
  & $\langle e_i,e_j,h_1\rangle$, $\langle f_i,f_j,h_1\rangle$ for $1\leq i<j\leq3$  & $\langle e_1,e_2,h_1\rangle$ \\
  & $\langle e_i,h_{5-i},h_1\rangle$, $\langle f_i,h_{5-i},h_1\rangle$ for $i=1,2,3$ & $\langle e_1,h_1,h_4\rangle$ \\
  & $\langle e_i,e_j,h_{i+j-1}\rangle$, $\langle f_i,f_j,h_{i+j-1}\rangle$ for $1\leq i<j\leq3$  & $\langle e_1,e_2,h_2\rangle$ \\
  & $\langle h_2,h_3,h_4\rangle$ &  $\langle h_2,h_3,h_4\rangle$ \\ \hline
4D   & $\langle e_1,e_2,e_3,h_1\rangle$, $\langle f_1,f_2,f_3,h_1\rangle$  & $\langle e_1,e_2,e_3,h_1\rangle$ \\
  & $\langle e_1,e_2,e_3,h_i\rangle$, $\langle f_1,f_2,f_3,h_i\rangle$ for $i=2,3,4$ & $\langle e_1,e_2,e_3,h_2\rangle$ \\
  & $\langle e_i,h_1,f_i\rangle\oplus\langle h_{5-i}\rangle$ for $i=1,2,3$  & $\langle e_1,h_1,f_1\rangle\oplus\langle h_4\rangle$  \\
  & $\langle e_i,e_j,h_{i+j-1},h_1\rangle$, $\langle f_i,f_j,h_{i+j-1},h_1\rangle$ for $1\leq i<j\leq3$ & $\langle e_1,e_2,h_1,h_2\rangle$ \\
  & $\langle h_2,h_3,h_4\rangle\oplus\langle h_1\rangle$ & $\langle h_2,h_3,h_4\rangle \oplus \langle h_1\rangle$ \\  \hline
\end{tabular}
\end{center}
\end{table}

Table \ref{Tab:SubAlg} shows the list of algebras.  We only need to consider one representative algebra from each of these classes, since other members are related via the involutions of Table \ref{Tab:Invg1234}. The chosen representative is shown in the final column.

For each representative case of the subalgebras listed in Table \ref{Tab:SubAlg}, we solve the equations $\{K_i,H\}=0$ for the function $\varphi(q_1,q_2,q_3)$ of (\ref{3d-gen}).  It often happens that the resulting Hamiltonian has a larger algebra of isometries, as indicated in Table \ref{Tab:SubAlg-Ham}.

\begin{table}[h]
\begin{center}
\caption{Invariant Hamiltonians for subalgebras of $\mathfrak{g}$}\label{Tab:SubAlg-Ham}\vspace{3mm}
\begin{tabular}{|l|l|l|l|}
\hline
 Dimension &             Representative             & $\varphi$ of Invariant Hamiltonian  & Maximal Isometry Algebra  \\[.10cm]\hline
2D  & $\langle e_1,h_1\rangle$  & $\varphi=q_2^2\psi\left(\frac{q_3}{q_2}\right)$ & $\langle e_1,h_1,f_1\rangle$ \\
   &     $\langle e_1,e_2\rangle$ & $\varphi=\psi(q_3)$ & $\langle e_1,e_2,h_2\rangle$ \\
 &  $\langle e_1,h_{4}\rangle$ & $\varphi=\psi\left(q_2^2+q_3^2\right)$ & $\langle e_1,h_{4}\rangle$  \\
  &    $\langle h_1,h_2\rangle$  & $\varphi=(q_1^2+q_2^2)\psi\left(\frac{q_3^2}{q_1^2+q_2^2}\right)$  &  $\langle h_1,h_2\rangle$ \\ \hline
3D   & $\langle e_1,e_2,e_3\rangle$  & $\varphi=1$  & $\langle e_1,e_2,e_3,h_2,h_3,h_4\rangle$  \\
 &  $\langle e_1,h_1,f_1\rangle$  & $\varphi=q_2^2\psi\left(\frac{q_3}{q_2}\right)$  &  $\langle e_1,h_1,f_1\rangle$  \\
  &   $\langle e_1,e_2,h_1\rangle$  & $\varphi=q_3^2$  & $\langle e_1,e_2,h_1,f_1,f_2,h_2\rangle$  \\
 & $\langle e_1,h_1,h_4\rangle$  & $\varphi=q_2^2+q_3^2$  & $\langle e_1,h_1,f_1\rangle\oplus\langle h_4\rangle$ \\
  &  $\langle e_1,e_2,h_2\rangle$ & $\varphi=\psi(q_3)$  &  $\langle e_1,e_2,h_2\rangle$  \\
  & $\langle h_2,h_3,h_4\rangle$   & $\varphi=\psi(q_1^2+q_2^2+q_3^2)$  &  $\langle h_2,h_3,h_4\rangle$ \\ \hline
 4D  &  $\langle e_1,e_2,e_3,h_1\rangle$ & There are {\em no} solutions.  & \\
 & $\langle e_1,e_2,e_3,h_2\rangle$ &  $\varphi=1$  & $\langle e_1,e_2,e_3,h_2,h_3,h_4\rangle$ \\
 & $\langle e_1,h_1,f_1\rangle\oplus\langle h_4\rangle$ &  $\varphi=q_2^2+q_3^2$  &  $\langle e_1,h_1,f_1\rangle\oplus\langle h_4\rangle$ \\
 & $\langle e_1,e_2,h_1,h_2\rangle$ & $\varphi=q_3^2$  & $\langle e_1,e_2,h_1,f_1,f_2,h_2\rangle$ \\
 &  $\langle h_2,h_3,h_4\rangle \oplus \langle h_1\rangle$ & $\varphi=q_1^2+q_2^2+q_3^2$  &  $\langle h_2,h_3,h_4\rangle \oplus \langle h_1\rangle$ \\  \hline
\end{tabular}
\end{center}
\end{table}
As can be seen, there are two cases with {\em genuinely} 2D algebras, with the solution depending upon an arbitrary function, $\psi$, of a single variable.
For the three genuinely 3 dimensional cases, the solution depends upon an arbitrary function, $\psi$, of a single variable.
With a 4D isometry algebra, the form of $\varphi$ is {\em explicitly} fixed.  There are two genuine cases.

\section{Geodesic Flows in 3D with Linear and Quadratic Integrals}\label{quadrat-int}

In later sections we seek particular cases of the Hamiltonian functions of Table \ref{Tab:SubAlg-Ham} which admit quadratic integrals of the type described below.  We use the method introduced in \cite{f19-3}, and used in \cite{f19-2}, to construct {\em quadratic invariants} out of conformal invariants.

A {\em quadratic conformal invariant} is any expression of the form
\begin{subequations}
\be\label{gen-quad}
F = \sum_{i,j=1}^{10} \beta_{ij} X_i X_j + \sigma(q_1,q_2,q_3) H,
\ee
where $\beta_{ij}$ are the ({\em constant}) coefficients of a symmetric matrix, $X_i$ are {\em linear} conformal invariants, and $\sigma(q_1,q_2,q_3)$ is an arbitrary function, which satisfies
$$
\{F,H\} = \left(\sum_{i=1}^3 \mu_i(q_1,q_2,q_3) p_i\right) H,
$$
where $\mu_i(q_1,q_2,q_3)$ are some functions.

We can ask whether there is a choice of coefficients $\beta_{ij}$ and $\sigma(q_1,q_2,q_3)$ for which $\mu_i(q_1,q_2,q_3)\equiv 0$, in which case {\em $F$ is a quadratic invariant}.

In fact, we will restrict our {\em first} quadratic integral to have the simpler form
\be\label{xij-quad}
F_1 = X_i X_j + \sigma(q_1,q_2,q_3) H,\quad\mbox{with}\quad \{F_1,H\}=0,
\ee
\end{subequations}
for some choice $1\leq i,j \leq 10$.  Since we have already restricted $\varphi$ of (\ref{3d-gen}) to satisfy one of the symmetry constraints, the single equation (\ref{xij-quad}) is enough to determine the functions $\psi$ (of the Table \ref{Tab:SubAlg-Ham}) and $\sigma$.  From $F_1$ we can then generate a Poisson algebra of integrals by taking Poisson brackets with the isometries $K_i$, along with the action of any involutions which preserve the isometry algebra {\em and} the form of $\varphi$.  We must also calculate the quantities $\{F_i,F_j\}$, which may necessitate additional integrals, after which we would need to calculate further Poisson brackets, and so on.  However, for each of the isometry algebras listed in Table \ref{Tab:SubAlg-Ham}, the resulting algebras are quite small, with simple Poisson relations.  Only occasionally does an additional integral appear when calculating $\{F_i,F_j\}$.

For a given isometry algebra we can systematically work through all choices of $i, j$ in (\ref{xij-quad}) and determine the list of compatible Hamiltonians (\ref{3d-gen}).  Again the action of $\{X_i X_j, K_\ell\}$, and of the appropriate involution(s), simplifies this calculation.

\subsection{Reductions to Two Dimensions}\label{sec:reductions}

In Sections \ref{2D-super} and \ref{3D-super} we present restrictions of Hamiltonians given in Table \ref{Tab:SubAlg-Ham}, for which an integral (\ref{xij-quad}) exists, and derive the corresponding Poisson algebras.

In Sections \ref{2D-super-reduce} to \ref{4D-super-reduce} we use the isometries to {\em reduce} our systems from 3 to 2 degrees of freedom.  We use the particular method of reduction introduced in \cite{f18-2}, referred to as the ``Kaluza-Klein'' reduction, since it is essentially the reverse procedure to the Kaluza-Klein \underline{extension}.  By adapting coordinates to a linear first integral, we can reduce from 3 to 2 degrees of freedom.  In principle, the lower dimensional system would possess {\em vector potential} terms, but in all our examples these can be removed by gauge transformation.

This approach was used in \cite{f19-2}, where it was shown that several superintegrable systems in 3 degrees of freedom can be reduced to Darboux-Koenigs systems in 2 degrees of freedom, but with the addition of a scalar potential function.  In the present paper we present {\em universal} reductions associated with each of the isometry algebras. For each of the (nontrivial) general Hamiltonians given in Table \ref{Tab:SubAlg-Ham} we give two reductions, which are generally {\em Liouville integrable}.  Any integral which commutes with the particular isometry, associated with the reduction, can also be reduced.  This gives rise to superintegrable reductions of Darboux-Koenigs type (\underline{one} linear and a quadratic integral).

Each reduction is associated with one particular isometry. It can be seen in Table \ref{Tab:g1234} that each conformal symmetry commutes with exactly 3 others.  What is not so apparent is that we can also find 3 other {\em quadratic} expressions which commute with this isometry, and that these 6 elements can be used to build the 6 conformal symmetries of the reduced space.  Our reduced quadratic integrals are then written in terms of this reduced conformal algebra.

\section{Superintegrable Restrictions with 2D Isometry Algebras}\label{2D-super}

In this section, we consider the kinetic energies $H$, with $\varphi$ given in Table \ref{Tab:SubAlg-Ham}, corresponding to the two nontrivial subalgebras $\left<e_1,h_4\right>$ and $\left<h_1,h_2\right>$, and list the cases consistent with quadratic integrals of the form (\ref{xij-quad}).  There are many choices of quadratic integrals for a given example, but in all our examples, the resulting algebra has rank 4, meaning that we only obtain 4 {\em functionally independent} integrals, $H, K_1, K_2, F_1$, so the corresponding systems are {\em superintegrable}, but not maximally so.  We can always choose $F_1$ in such a way that we only need to introduce one additional integral $F_2$.

We find that, in each case, {\em both} functions $F_i$ {\em commute} with one particular isometry (say $K_1$).  This follows from one of two mechanisms:
\bea
&&     \{K_1,F_1\}=0,\;\;\; \{K_2,F_1\}= c F_2,\;\;\;  \{K_1,K_2\}=0, \quad\Rightarrow\quad   \{K_1,F_2\}=0,    \nn\\
&&       \{K_1,F_1\}=0,\;\;\; \iota_{ab} K_1=\pm K_1,  \;\;\; \iota_{ab} F_1= c F_2    \quad\Rightarrow\quad   \{K_1,F_2\}=0,   \nn
\eea
where $c$ is a constant and $\iota_{ab}$ denotes the relevant involution.

When we consider reductions to 2 dimensional systems in Section \ref{2D-super-reduce}, we use one particular isometry for the reduction and any commuting integral will reduce to the 2 dimensional space.  We then find that one of our reductions has one isometry and one or two quadratic integrals, so is of Darboux-Koenigs type, whilst the other reduction is to a system with an isometry, but {\em no} independent quadratic integral.

\subsection{Systems with Isometry Algebra $\left<e_1,h_4\right>$}\label{2D-super-e1h4}

This algebra is {\em commutative} and invariant under the action of $\iota_{23}$, as is the general Hamiltonian with conformal factor $\varphi=\psi\left(q_2^2+q_3^2\right)$.  We are led to three cases of the function $\psi\left(q_2^2+q_3^2\right)$, which we list below.

There are more possibilities, but they all lead to systems with larger isometry algebras.

\subsubsection{The Case $\psi(z)=\frac{z}{\alpha z+\beta}$}\label{sec:e1h4eq1}

The Hamiltonian
\begin{subequations}
\be\label{e1h4eq1}
  H=\frac{q_2^2+q_3^2}{\alpha(q_2^2+q_3^2)+\beta}\, \left(p_1^2+p_2^2+p_3^2\right),
\ee
is derived from quadratic integrals of the form
\be\label{e1h4eq1-F1F2}
  F_1=h_1^2-4\alpha \left(q_1^2+q_2^2+q_3^2\right) H,\quad F_2=e_1h_1+2\alpha q_1 H.
\ee
The integrals $H,e_1,h_4,F_1,F_2$ form a Poisson algebra with non-zero Poisson brackets given by
\be\label{e1h4eq1-pb}
\{e_1,F_1\}=4 F_2,\quad \{e_1,F_2\} = 2e_1^2-2\alpha H,\quad \{F_1,F_2\} = -4e_1F_1.
\ee
Only four of these functions are independent, since they obey the algebraic constraint:
\be\label{e1h4eq1-constraint}
  F_2^2 = \left(e_1^2-\alpha H\right) F_1+ \frac{1}{4}\alpha H \left(16 \beta H-h_4^2\right).
\ee
\end{subequations}
These integrals are \underline{invariant} under the action of $\iota_{23}$.

\subsubsection{The Case $\psi(z)=\frac{1}{\alpha z+\beta}$}\label{sec:e1h4eq2}

The Hamiltonian
\begin{subequations}
\be\label{e1h4eq2}
  H=\frac{1}{\alpha(q_2^2+q_3^2)+\beta}\left(p_1^2+p_2^2+p_3^2\right),
\ee
is derived from quadratic integrals of the form
\be\label{e1h4eq2-F1F2}
  F_1=e_2e_3-\alpha q_2q_3 H,\quad F_2=e_2^2-e_3^2+\alpha (q_3^2-q_2^2)H.
\ee
The integrals $H,e_1,h_4,F_1,F_2$ form a Poisson algebra with non-zero Poisson brackets given by
\be\label{e1h4eq2-pb}
\{h_4,F_1\}=4 F_2,\quad \{h_4,F_2\} = -16F_1,\quad \{F_1,F_2\} = -\alpha h_4H.
\ee
Only four of these functions are independent, since they obey the algebraic constraints:
\be\label{e1h4eq2-constraint}
  16F_1^2+4F_2^2-4(e_1^2-\beta H)^2-\alpha h_4^2 H=0.
\ee
\end{subequations}

The action of the involution is summarised in $\iota_{23}:\ (H,e_1,h_4,F_1,F_2)\mapsto(H,e_1,-h_4,F_1,-F_2)$.

\subsubsection{The Case $\psi(z)=\frac{\sqrt{z}}{\alpha \sqrt{z}+\beta}$}\label{sec:e1h4eq3}

The Hamiltonian
\begin{subequations}
\be\label{e1h4eq3}
  H=\frac{\sqrt{q_2^2+q_3^2}}{\alpha\sqrt{q_2^2+q_3^2}+\beta} \left(p_1^2+p_2^2+p_3^2\right),
\ee
is derived from quadratic integrals of the form
\be\label{e1h4eq3-F1F2}
  F_1=e_2h_4-\frac{2\beta q_3}{\sqrt{q_2^2+q_3^2}}H,  \quad F_2=e_3h_4+\frac{2\beta q_2}{\sqrt{q_2^2+q_3^2}}H.
\ee

These integrals form the Poisson algebra with non-zero Poisson brackets given by
\be\label{e1h4eq3-pb}
\{h_4,F_1\}=-4 F_2,\quad \{h_4,F_2\} = 4F_1,\quad \{F_1,F_2\} = 4h_4(\alpha H-e_1^2),
\ee
and satisfy the constraint
\be\label{e1h4eq3-constraint}
  F_1^2+F_2^2+ h_4^2 \left(e_1^2-\alpha H\right)=4\beta^2 H^2.
\ee
\end{subequations}

The action of the involution is summarised in $\iota_{23}:\ (H,e_1,h_4,F_1,F_2)\mapsto(H,e_1,-h_4,-F_2,-F_1)$.

\subsection{Systems with Isometry Algebra $\left<h_1,h_2\right>$}\label{2D-super-h1h2}

This algebra is {\em commutative} and invariant under the action of both $\iota_{12}$ and $\iota_{ef}$, as is the general Hamiltonian with conformal factor $\varphi=(q_1^2+q_2^2)\psi\left(\frac{q_3^2}{q_1^2+q_2^2}\right)$.  We are led to four cases of the function $\psi\left(\frac{q_3^2}{q_1^2+q_2^2}\right)$, which we list below.

There are more possibilities, but they all lead to systems with larger isometry algebras.

\subsubsection{The Case $\psi(z)=\frac{(1+z) z}{\alpha +\beta z}$}\label{sec:h1h2eq1}

The Hamiltonian
\begin{subequations}
\be\label{h1h2eq1}
  H=\frac{\left(q_1^2+q_2^2+q_3^2\right)q_3^2}{\alpha\left(q_1^2+q_2^2\right)+\beta q_3^2}\left(p_1^2+p_2^2+p_3^2\right),
\ee
is derived from quadratic integrals of the form
\be\label{h1h2eq1-F1F2}
  F_1=h_3h_4+\frac{8\alpha q_1q_2}{q_3^2}H,\quad F_2 = 4 h_3^2-h_4^2 - 16\alpha\, \frac{q_1^2-q_2^2}{q_3^2}\, H.
\ee
The integrals $H,h_1,h_2,F_1,F_2$ form a Poisson algebra with non-zero Poisson brackets given by
\be\label{h1h2eq1-pb}
\{h_2,F_1\}= F_2,\quad \{h_2,F_2\} = -16 F_1,\quad \{F_1,F_2\} = -32h_2(h_1^2+h_2^2+4(2\alpha-\beta)H).
\ee
Only four of these functions are independent, since they obey the algebraic constraint:
\be\label{h1h2eq1-constraint}
   16 F_1^2+F_2^2+128\left(\beta h_1^2+(\beta-2\alpha)h_2^2-2 \beta^2 H\right)H-16\left(h_1^2+h_2^2\right)^2=0.
\ee
\end{subequations}
The actions of $\iota_{12}$ and $\iota_{ef}$ on integrals are summarised in
$$
\iota_{12}:\ (H,h_1,h_2,F_1,F_2)\mapsto(H,h_1,-h_2,F_1,-F_2), \quad \iota_{ef}:\ (H,h_1,h_2,F_1,F_2)\mapsto(H,-h_1,h_2,F_1,F_2).
$$

\subsubsection{The Case $\psi(z)=\frac{\sqrt{1+z}}{\alpha \sqrt{1+z} +\beta \sqrt{z}}$}\label{sec:h1h2eq2}

The Hamiltonian
\begin{subequations}
\be\label{h1h2eq2}
  H=\frac{\sqrt{q_1^2+q_2^2+q_3^2}\, \left(q_1^2+q_2^2\right)}{\alpha\sqrt{q_1^2+q_2^2+q_3^2}+\beta q_3}\, \left(p_1^2+p_2^2+p_3^2\right),
\ee
is derived from quadratic integrals of the form
\be\label{h1h2eq2-F1F2}
  F_1=h_1e_3-\frac{\beta}{\sqrt{q_1^2+q_2^2+q_3^2}}\, H,\quad F_2=h_1f_3+2\beta\, {\sqrt{q_1^2+q_2^2+q_3^2}}\, H.
\ee
The integrals $H,h_1,h_2,F_1,F_2$ form a Poisson algebra with non-zero Poisson brackets given by
\be\label{h1h2eq2-pb}
\{h_1,F_1\}= -2 F_1,\quad \{h_1,F_2\} = 2 F_2,\quad \{F_1,F_2\} = -2h_1(2h_1^2+h_2^2-4\alpha H).
\ee
Only four of these functions are independent, since they obey the algebraic constraint:
\be\label{h1h2eq2-constraint}
    2F_1F_2+4H\left(\beta^2H-\alpha h_1^2\right)+h_1^2(h_1^2+h_2^2)=0.
\ee
\end{subequations}
The actions of $\iota_{12}$ and $\iota_{ef}$ on integrals are summarised in
$$
\iota_{12}:\ (H,h_1,h_2,F_1,F_2)\mapsto(H,h_1,-h_2,F_1,F_2), \quad \iota_{ef}:\ (H,h_1,h_2,F_1,F_2)\mapsto\left(H,-h_1,h_2,\frac12 F_2,2F_1\right).
$$

\subsubsection{The Case $\psi(z)=\frac{1+z}{\alpha  +\beta \sqrt{z}}$}\label{sec:h1h2eq3}

The Hamiltonian
\begin{subequations}
\be\label{h1h2eq3}
  H=\frac{\sqrt{q_1^2+q_2^2}\, \left(q_1^2+q_2^2+q_3^2\right)}{\alpha\sqrt{q_1^2+q_2^2}+\beta q_3}\, \left(p_1^2+p_2^2+p_3^2\right),
\ee
is derived from quadratic integrals of the form
\be\label{h1h2eq3-F1F2}
   F_1=h_2h_3-\frac{2\beta q_2}{\sqrt{q_1^2+q_2^2}}H,\quad F_2=h_2h_4-\frac{4\beta q_1}{\sqrt{q_1^2+q_2^2}}H.
\ee
The integrals $H,h_1,h_2,F_1,F_2$ form a Poisson algebra with non-zero Poisson brackets given by
\be\label{h1h2eq3-pb}
\{h_2,F_1\}= -F_2,\quad \{h_2,F_2\} = 4 F_1,\quad \{F_1,F_2\} = 4h_2(4\alpha H-h_1^2-2h_2^2).
\ee
Only four of these functions are independent, since they obey the algebraic constraint:
\be\label{h1h2eq3-constraint}
    4F_1^2+F_2^2-16 H\left(\beta^2H+\alpha h_2^2\right)+4h_2^2(h_1^2+2h_2^2)=0.
\ee
\end{subequations}
The actions of $\iota_{12}$ and $\iota_{ef}$ on integrals are summarised in
$$
\iota_{12}:\ (H,h_1,h_2,F_1,F_2)\mapsto \left(H,h_1,-h_2,\frac12 F_2,2F_1\right), \quad \iota_{ef}:\ (H,h_1,h_2,F_1,F_2)\mapsto(H,-h_1,h_2,F_1,F_2).
$$

\subsubsection{The Case $\psi(z)=\frac{z}{\alpha +\beta z}$}\label{sec:h1h2eq4}

The Hamiltonian
\begin{subequations}
\be\label{h1h2eq4}
  H=\frac{\left(q_1^2+q_2^2\right)q_3^2}{\alpha\left(q_1^2+q_2^2\right)+\beta q_3^2}\, \left(p_1^2+p_2^2+p_3^2\right),
\ee
is derived from quadratic integrals of the form
\be\label{h1h2eq4-F1F2}
   F_1=e_3^2-\frac\alpha{q_3^2}\, H,\quad    F_2=f_3^2-\frac{4\alpha\left(q_1^2+q_2^2+q_3^2\right)^2}{q_3^2}\, H.
\ee
The integrals $H,h_1,h_2,F_1,F_2$ form a Poisson algebra with non-zero Poisson brackets given by
\be\label{h1h2eq4-pb}
\{h_1,F_1\}= -4 F_1,\quad \{h_1,F_2\} = 4 F_2,\quad \{F_1,F_2\} = 4h_1(h_1^2+h_2^2-4(\alpha+\beta)H).
\ee
Only four of these functions are independent, since they obey the algebraic constraint:
\be\label{h1h2eq4-constraint}
    F_1F_2+2\left((\alpha+\beta)h_1^2+(\beta-\alpha)h_2^2-2(\alpha-\beta)^2H\right)\, H-\frac14 \left(h_1^2+h_2^2\right)^2=0.
\ee
\end{subequations}
The actions of $\iota_{12}$ and $\iota_{ef}$ on integrals are summarised in
$$
\iota_{12}:\ (H,h_1,h_2,F_1,F_2)\mapsto(H,h_1,-h_2,F_1,F_2), \quad \iota_{ef}:\ (H,h_1,h_2,F_1,F_2)\mapsto\left(H,-h_1,h_2,\frac14 F_2,4F_1\right).
$$

\section{Superintegrable Restrictions with 3D Isometry Algebras}\label{3D-super}

Here we consider the three nontrivial 3D cases listed in Table \ref{Tab:SubAlg-Ham}.  Since these algebras are not commutative, they generate larger, more complex Poisson algebras from the given quadratic function $F_1$.

\subsection{Systems with Isometry Algebra $\left<e_1,h_1,f_1\right>$}\label{3D-super-e1h1f1}

As given in Table \ref{Tab:SubAlg-Ham}, the general conformal factor in this case is $\varphi=q_2^2\psi\left(\frac{q_3}{q_2}\right)$.  This isometry algebra is invariant under the action of both $\iota_{23}$ and $\iota_{ef}$, as is the {\em general form} of Hamiltonian with this conformal factor, although the specific form of $\psi$ changes in the case of $\iota_{23}$.  Depending upon our choice of quadratic integrals, the full Poisson algebra of integrals may or may not be invariant with respect to the involutions.  Up to equivalence under point transformations, we obtain two cases, both of which are invariant under the action of $\iota_{ef}$, but not $\iota_{23}$.  Since both of these were discussed in \cite{f19-2}, only brief details are presented here.

\subsubsection{The Case $\psi(z)=\frac{z^2}{\alpha  +\beta z^2}$}\label{sec:e1h1f1eq1}

The Hamiltonian
\begin{subequations}
\be\label{e1h1f1eq1}
  H=\frac{q_2^2q_3^2}{\alpha q_2^2+\beta q_3^2}\, \left(p_1^2+p_2^2+p_3^2\right),
\ee
is derived from a quadratic integral of the form
\be\label{e1h1f1eq1-F1}
  F_1=e_2f_2-\sigma(q_1,q_2,q_3)H,\quad\mbox{where}\quad \sigma = 2\beta\, \frac{q_1^2+q_3^2}{q_2^2}.
\ee
\end{subequations}
The Poisson bracket of $F_1$ with the isometries, leads to a further four quadratic integrals $F_2, \dots , F_5$, giving us an 8 dimensional Poisson algebra of first integrals of $H$.  The Poisson relations are given in Table 6 of \cite{f19-2}.  Since $H, e_1, {\cal C}$ (where ${\cal C} = e_1 f_1+\frac{1}{4}\, h_1^2$ is the Casimir function) are in involution, the system is Liouville integrable.  Furthermore, since $H,\, e_1,\, h_1,\, f_1$ and $F_1$ are independent, the system is maximally superintegrable.

\subsubsection{The Case $\psi(z)=\frac{\sqrt{1+z^2}}{\alpha \sqrt{1+z^2}  +\beta z}$}\label{sec:e1h1f1eq2}

The Hamiltonian
\begin{subequations}
\be\label{e1h1f1eq2}
   H=\frac{q_2^2\sqrt{q_2^2+q_3^2}}{\alpha\sqrt{q_2^2+q_3^2}+\beta q_3}\, \left(p_1^2+p_2^2+p_3^2\right),
\ee
is derived from a quadratic integral of the form
\be\label{e1h1f1eq2-F1}
  F_1= \frac{1}{2}\, e_2 h_4-\sigma(q_1,q_2,q_3)H,\quad\mbox{where}\quad \sigma = \frac{\beta \left(q_2^2+2 q_3^2\right)+2\alpha q_3 \sqrt{q_2^2+q_3^2}}{q_2^2 \sqrt{q_2^2+q_3^2}} .
\ee
\end{subequations}
The Poisson bracket of $F_1$ with the isometries, together with the bracket $\{F_1,F_2\}$, lead to a further three quadratic integrals $F_2, \dots , F_4$, giving us an 7 dimensional Poisson algebra of first integrals of $H$.  The Poisson relations are given in Table 8 of \cite{f19-2}.  Since $H, e_1, {\cal C}$ (where ${\cal C} = e_1 f_1+\frac{1}{4}\, h_1^2$ is the Casimir function) are in involution, the system is Liouville integrable.  Furthermore, since $H,\, e_1,\, h_1,\, f_1$ and $F_1$ are independent, the system is maximally superintegrable.

\br[Comparison with \cite{f19-2}]\label{e1h1f1eq2-rem}
In \cite{f19-2} we discuss the equivalent system with algebra $\left<e_3,h_1,f_3\right>$, so to compare formula we must transform all formulae in accordance with $\iota_{13}$.
\er

\subsection{Systems with Isometry Algebra $\left<e_1,e_2,h_2\right>$}\label{3D-super-e1e2h2}

As given in Table \ref{Tab:SubAlg-Ham}, the general conformal factor in this case is $\varphi=\psi(q_3)$.  This isometry algebra and general Hamiltonian are invariant under the action of $\iota_{12}$, which helps us build the Poisson algebra of integrals.  Up to equivalence under point transformations, we obtain two cases.

\subsubsection{The Case $\psi(z)=\frac1{\alpha z+\beta}$}\label{sec:e1e2h2eq1}

The Hamiltonian
\be\label{e1e2h2eq1}
  H=\frac1{\alpha q_3+\beta}\, \left(p_1^2+p_2^2+p_3^2\right),
\ee
admits the quadratic integrals
\bea
&&  F_1=e_1h_3-\frac\alpha2 q_1^2H,\quad F_2=e_1e_3-\frac\alpha2 q_1H,\quad F_3=-e_1h_4+2e_2h_3-2\alpha q_1q_2H, \nn \\[-2mm]
&&                        \label{e1e2h2eq1-Fi}   \\[-2mm]
&&  F_4=e_2e_3-\frac\alpha2 q_2H,\quad F_5=e_2h_4+\alpha q_2^2H.   \nn
\eea
They form a Poisson algebra of rank 5, satisfying the relations given in Table \ref{Tab:e1e2h2}.

\begin{table}[h]\caption{The Poisson algebra of first integrals of (\ref{e1e2h2eq1})}\label{Tab:e1e2h2}
\begin{center}
\resizebox{\textwidth}{!}{
\renewcommand\arraystretch{1.26}\begin{tabular}{|c||c|c|c||c|c|c|c||c|}
\hline
       &$e_1$    &$e_2$     &$h_2$   &$F_1$   &$F_2$    &$F_3$  &$F_4$ &$F_5$\\[1.5mm]\hline\hline
$e_1$  &$0$ &$0$ &$-2e_2$  &$-2F_2$  &$\frac\alpha2 H$ &$-4F_4$  &$0$ &$0$\\[1.5mm]\hline
$e_2$  &    &$0$ &$2e_1$   &$0$   &$0$  &$-4F_2$    &$\frac\alpha2 H$ &$4F_4$\\[1.5mm]\hline
$h_2$  &    &    &$0$     &$F_3$  &$2F_4$   &$-4(2F_1+F_5)$  &$-2F_2$ &$2F_3$\\[1.5mm]\hline\hline
$F_1$ &    &  &  &$0$ &$-2e_1(2e_1^2+e_2^2-\beta H)$ &$4h_2(2e_1^2+e_2^2-\beta H)$ &$-2e_1^2e_2$ &$-4e_1e_2h_2$\\[1.5mm]\hline
$F_2$ &    &  &  &   &$0$ &$4e_2(3e_1^2+e_2^2-\beta H)$ &$0$ &$-4e_1e_2^2$  \\[1.5mm]\hline
$F_3$ & &  &  &  &   &$0$ &$-4e_1(e_1^2+3e_2^2-\beta H)$ &$-8h_2(e_1^2+2e_2^2-\beta H)$\\[1.5mm]\hline
$F_4$ & & &  &  &  &   &$0$ &$-4e_2(e_1^2+2e_2^2-\beta H)$ \\[1.5mm]\hline
$F_5$ & & &  &  &  &   &    &$0$\\[1.5mm]\hline
\end{tabular}
}
\end{center}
\end{table}

The actions of $\iota_{12}$ on the integrals are summarized in
$$
\iota_{12} : \left(H,e_1,e_2,h_2,F_1,F_2,F_3,F_4,F_5\right) \mapsto \left(H,e_2,e_1,-h_2,-\frac12 F_5,F_4,F_3,F_2,-2F_1\right).
$$
This algebra satisfies the following four constraints:
\bea
&&  e_1F_4-e_2F_2-\frac\alpha4 h_2 H=0,\quad e_1F_3-4e_2F_1+2h_2F_2=0,  \nn\\
&&    2e_1F_5+e_2F_3-2h_2F_4=0,\quad e_1e_2(e_1^2+e_2^2-\beta H)+F_2F_4+\frac\alpha8 F_3H=0.  \nn
\eea

\br
This Hamiltonian is related to $H=\frac1{\alpha q_2 +\beta q_3+\gamma}\, \left(p_1^2+p_2^2+p_3^2\right)$ of Sec. 4.2 in \cite{f19-2}, through a rotation in the $(q_2,q_3)$ plane.  The Poisson relations of Table \ref{Tab:e1e2h2} are simpler than those of Table 7 in \cite{f19-2}.
\er

\subsubsection{The Case $\psi(z)=\frac{z^2}{\alpha z^2+\beta}$}\label{sec:e1e2h2eq2}

The Hamiltonian
\be\label{e1e2h2eq2}
  H=\frac{q_3^2}{\alpha q_3^2+\beta}\,\left(p_1^2+p_2^2+p_3^2\right),
\ee
is compatible with the quadratic integral
\be\label{e1e2h2eq2-F1}
F_1 = e_3 f_3 + 2 \left(\alpha q_3^2- \beta \left(\frac{q_1^2+q_2^2}{q_3^2}\right)\right)\, H,
\ee
which (taking Poisson brackets with the isometries) generates another three quadratic integrals.  The resulting Poisson algebra is equivalent (under $\iota_{23}$ and a small change of notation) to that shown in Table 3 of \cite{f19-2}.  The algebra has rank 5 and defines a maximally superintegrable system.

\subsection{Systems with Isometry Algebra $\left<h_2,h_3,h_4\right>$}\label{3D-super-h2h3h4}

As given in Table \ref{Tab:SubAlg-Ham}, the general conformal factor in this case is $\varphi=\psi\left(q_1^2+q_2^2+q_3^2\right)$.  This isometry algebra is invariant under the action of \underline{all} the involutions of Table \ref{Tab:Invg1234}.  The general Hamiltonian is invariant under 3 of the involutions, but \underline{not} $\iota_{ef}$.  Invariance under $\iota_{ef}$ implies that $\varphi=q_1^2+q_2^2+q_3^2$, which has an additional isometry, $h_1$, and is discussed in Section \ref{4D-super-reduce}.  The symmetry of the Poisson algebra of integrals, under involutions, depends upon the choice of quadratic integrals.  Up to equivalence under point transformations, we obtain two cases.

\subsubsection{The Case $\psi(z)=\frac{\sqrt{z}}{\alpha \sqrt{z} +\beta}$}\label{sec:h2h3h4eq1}

The Hamiltonian
\be\label{h2h3h4eq1}
  H=\frac{\sqrt{q_1^2+q_2^2+q_3^2}}{\alpha\sqrt{q_1^2+q_2^2+q_3^2}+\beta}\, \left(p_1^2+p_2^2+p_3^2\right),
\ee
has the three quadratic integrals
\begin{subequations}
\be\label{h2h3h4eq1-Fi}
  F_i=e_ih_1+\frac{q_i\left(2\alpha\sqrt{q_1^2+q_2^2+q_3^2}+\beta\right)}{\sqrt{q_1^2+q_2^2+q_3^2}}H,\qquad i=1,2,3.
\ee
These integrals form a Poisson algebra with the Poisson relations given in Table \ref{Tab:h2h3h4:1} and satisfy the constraints
\be\label{h2h3h4eq1-cons}
  h_4F_1+2h_3F_2-2h_2F_3=0,\quad F_1^2+F_2^2+F_3^2-\beta^2H^2-\alpha (h_2^2+h_3^2+\frac14h_4^2)H=0.
\ee
\end{subequations}

\begin{table}[h]
\begin{center}
\caption{The Poisson algebra of integrals of (\ref{h2h3h4eq1})}\label{Tab:h2h3h4:1}\vspace{3mm}
\begin{tabular}{c|ccc|ccc}
\hline
     &$h_2$    &$h_3$  &$h_4$     &$F_1$   &$F_2$   &$F_3$\\[.10cm]\hline
$h_2$ &$0$    &$-h_4$  &$4h_3$    &$2F_2$    &$-2F_1$     &$0$\\
$h_3$ &       &$0$     &$-4h_2$   &$2F_3$    &$0$  &$-2F_1$\\
$h_4$ &&               &$0$       &$0$       &$-4F_3$  &$4F_2$\\\hline
$F_1$ &&&     &$0$        &$-2\alpha h_2H$ &$-2\alpha h_3H$ \\
$F_2$ &&&&    &$0$        &$\alpha h_4H$       \\
$F_3$ &&&&&   &$0$\\
\hline
\end{tabular}
\end{center}
\end{table}

The actions of the involutions on the integrals are summarized in Table \ref{Tab:h2h3h4:2}.
\begin{table}[h]
\begin{center}
\caption{The actions of the involutions on (\ref{h2h3h4eq1}) and its integrals} \label{Tab:h2h3h4:2} \vspace{3mm}
{\begin{tabular}{c|c|ccc|ccc}
\hline
             & $H$ &$h_2$   &$h_3$      &$h_4$      &$F_1$ &$F_2$ &$F_3$ \\[.10cm]\hline
$\iota_{12}$ & $H$ &$-h_2$ &$-\frac12h_4$ &$-2h_3$  &$F_2$ &$F_1$ &$F_3$\\[1mm]\hline
$\iota_{13}$ & $H$ &$\frac12h_4$ &$-h_3$ &$2h_2$  &$F_3$ &$F_2$ &$F_1$\\[1mm]\hline
$\iota_{23}$ & $H$ &$h_3$ &$h_2$ &$-h_4$  &$F_1$ &$F_3$ &$F_2$\\[1mm]\hline
\end{tabular}}
\end{center}
\end{table}

\subsubsection{The Case $\psi(z)=\frac1{\alpha z +\beta}$}\label{sec:h2h3h4eq2}

The Hamiltonian
\be\label{h2h3h4eq2}
  H=\frac{1}{\alpha(q_1^2+q_2^2+q_3^2)+\beta}\, \left(p_1^2+p_2^2+p_3^2\right),
\ee
has the quadratic integral $F_1 = e_1^2 -\alpha q_1^2 H$, which generates a further five quadratic integrals.  These 10 integrals satisfy Poisson relations equivalent to those given in Table 10 of \cite{f19-2}, which is invariant under the action of $\iota_{23}$.

\section{Reduction to 2 Dimensional Systems: 2D Isometry Algebras}\label{2D-super-reduce}

In this section we use the approach described in Section \ref{sec:reductions} to reduce our systems of Section \ref{2D-super} to 2 dimensional spaces.  Since the two isometries {\em commute}, we can {\em simultaneously} adapt coordinates and then adjust the canonical transformation so that the resulting Hamiltonian takes ``conformal form'' in the remaining 2 dimensions.  Any integrals which {\em commute} with this isometry will reduce to give an integral in the two dimensional space, which can be expressed in terms of the reduced conformal algebra.

Since all our systems have integrals that commute with {\em just one} of the isometries, we obtain two types of reduction: systems of Darboux-Koenigs type, when the quadratic integrals can reduce, or systems with just one 2D isometry and no quadratic integrals.

\subsection{Systems with Isometry Algebra $\left<e_1,h_4\right>$}\label{2D-super-reduce-e1h4}

We give a pair of {\em universal} reductions of the general Hamiltonian in this class:
\be\label{3d-e1h4}
H = \psi\left(q_2^2+q_3^2\right) \left(p_1^2+p_2^2+p_3^2\right).
\ee
We then give the reductions of the three cases presented in Section \ref{2D-super-e1h4}.

Since the isometries {\em commute}, we can {\em simultaneously} reduce them to $(e_1,h_4)=(P_1,P_3)$, giving us the form of $Q_1, Q_3$, up to arbitrary functions of $q_2^2+q_3^2$, which is the common invariant of $e_1$ and $h_4$.  Since
$$
 \{q_1,e_1\}=\left\{\frac14\arctan\left(\frac{q_2}{q_3}\right),h_4\right\}=1,
$$
the most general canonical transformation which achieves this is:
\be\label{e1h4-genS}
S = (q_1+w_1(z)) P_1+w_2(z)P_2+\left(\frac{1}{4}\arctan\left(\frac{q_2}{q_3}\right)+w_3(z)\right)P_3,\quad\mbox{where}\quad  z=q_2^2+q_3^2.
\ee
We first require that the coefficients of $P_iP_j$ are zero, giving
$$
w_1'(z)w_2'(z)=w_2'(z)w_3'(z)=w_1'(z)w_3'(z)=0 \quad\Rightarrow\quad   w_1'(z)  =  w_3'(z)  =  0,
$$
since $w_2'(z)$ {\em cannot} be zero.  We can therefore take $w_1(z)=w_3(z)=0$.   The variable $Q_2$ is then chosen to equate coefficients of $P_1^2,P_2^2$ or of $P_2^2,P_3^2$, giving an ODE for $w_2(z)$ in each case, leading to the two transformations.

\subsubsection*{Reduction with respect to $e_1$}

Here we have the generating function
\begin{subequations}
\be\label{TRe1h4:1}
S = q_1P_1+\frac18\log\left(q_2^2+q_3^2\right)\, P_2+\frac14\arctan\left(\frac{q_2}{q_3}\right)P_3,
\ee
giving the Hamiltonian
\be  \label{2d-e1h4:1}
 H = \frac{1}{16}\, e^{-8 Q_2}\, \psi\left(e^{8 Q_2}\right) \left(P_2^2+P_3^2+ 16 e^{8 Q_2} P_1^2\right).
\ee
We can think of this as corresponding to a conformally flat metric in the $2-3$ space, defined by $P_1 = \mbox{const.}$, with $16 e^{8 Q_2} P_1^2$ (times the conformal factor) corresponding to a potential.  Since the conformal factor is a function of only $Q_2$, the momentum $P_3$ corresponds to a Killing vector (in 2D), and, in this case, is a first integral of the entire Hamiltonian.

As remarked in Section \ref{sec:reductions}, we can build the 6 dimensional conformal algebra for this 2D metric (kinetic energy).  First note that $e_2,\, e_3,\, h_4,\, e_1f_1+\frac{1}{4} h_1^2,\, e_2 f_2-\frac{1}{2} h_2^2,\,  e_3 f_3-\frac{1}{2} h_3^2$ {\em commute} with $e_1$.  Writing these in terms of $Q_i,P_i$ and discarding the $P_1^2$ components in the quadratic ones, we can derive the following 6 conformal elements:
\bea
&&  {\cal T}_{e_1} = \frac{1}{4} e^{4Q_2} (P_2 \sin 4Q_3-P_3 \cos 4Q_3),\;\;\; {\cal T}_{h_1} = \frac{1}{2} P_2,\;\;\; {\cal T}_{f_1}= -\frac{1}{4} e^{-4Q_2} (P_2 \sin 4Q_3+P_3 \cos 4Q_3), \nn   \\[-1mm]
&&    \label{e1h4-conf-e1}    \\[-1mm]
&&  {\cal T}_{e_2} = \frac{1}{4} e^{4Q_2} (P_2 \cos 4Q_3+P_3 \sin 4Q_3),\;\;\; {\cal T}_{h_2} = -\frac{1}{2} P_3,\;\;\; {\cal T}_{f_2}= \frac{1}{2} e^{-4Q_2} (P_3 \sin 4Q_3-P_2 \cos 4Q_3),  \nn
\eea
which satisfy the relations of $\mathfrak{g}_1 + \mathfrak{g}_2$ in Table \ref{Tab:g1234}, together with the algebraic constraints:
\be\label{e1h4-conf-constraints}
{\cal T}_{e_1}{\cal T}_{f_1}+\frac{1}{4} {\cal T}_{h_1}^2+\frac{1}{2} \left({\cal T}_{e_2}{\cal T}_{f_2}-\frac{1}{2} {\cal T}_{h_2}^2\right)=0,
                          \quad {\cal T}_{e_1}{\cal T}_{f_2}-2 {\cal T}_{e_2} {\cal T}_{f_1} +{\cal T}_{h_1}{\cal T}_{h_2} =0.
\ee
\end{subequations}

\subsubsection*{Reduction with respect to $h_4$}

Here we have the generating function
\begin{subequations}
\be\label{TRe1h4:2}
S = q_1P_1+\sqrt{q_2^2+q_3^2}\, P_2+\frac14\arctan\left(\frac{q_2}{q_3}\right)P_3,
\ee
giving the Hamiltonian
\be  \label{2d-e1h4:2}
 H = \psi\left(Q_2^2\right) \left(P_1^2+P_2^2+\frac{P_3^2}{16 Q_2^2}\right).
\ee
This case corresponds to a conformally flat metric in the $1-2$ space, defined by $P_3 = \mbox{const.}$, with $\frac{P_3^2}{16 Q_2^2}$ (times the conformal factor) corresponding to a potential.  Since the conformal factor is a function of only $Q_2$, the momentum $P_1$ corresponds to a Killing vector (in 2D), and, again, is a first integral of the entire Hamiltonian.

Again, we can build the 6 dimensional conformal algebra for this 2D metric (kinetic energy).  First note that $e_1,\, h_1,\, f_1,\, e_2^2+e_3^2,\, h_2^2+h_3^2,\,  f_2^2+f_3^2$ {\em commute} with $h_4$.
Writing these in terms of $Q_i,P_i$ and discarding the $P_3^2$ components in the quadratic ones, we can derive the following 6 conformal elements:
\bea
&&  {\cal T}_{e_1} = P_1,\;\;\; {\cal T}_{h_1} = -2(Q_1P_1+Q_2P_2),\;\;\; {\cal T}_{f_1}= (Q_2^2-Q_1^2) P_1-2 Q_1Q_2 P_2,  \nn  \\[-1mm]
&&    \label{e1h4-conf-h4}    \\[-1mm]
&&  {\cal T}_{e_2} = P_2,\quad {\cal T}_{h_2} = 2 (Q_1P_2-Q_2 P_1),\quad {\cal T}_{f_2}= 2(Q_1^2-Q_2^2) P_2-4 Q_1Q_2 P_1, \nn
\eea
\end{subequations}
which satisfy the relations of $\mathfrak{g}_1 + \mathfrak{g}_2$ in Table \ref{Tab:g1234}, together with the algebraic constraints (\ref{e1h4-conf-constraints}).

\subsubsection{The Case $\psi(z)=\frac{z}{\alpha z+\beta}$}\label{sec:e1h4eq1-red}

With
\begin{subequations}
\be\label{e1h4eq1-red}
  H=\frac{q_2^2+q_3^2}{\alpha(q_2^2+q_3^2)+\beta}\, \left(p_1^2+p_2^2+p_3^2\right),
\ee
we find that the Hamiltonians (\ref{2d-e1h4:1}) and (\ref{2d-e1h4:2}) give
\bea
\tilde{H} &=& \frac1{16(\alpha \text{e}^{8Q_2}+\beta)}\, \left(P_2^2+P_3^2+16 \text{e}^{8Q_2}P_1^2\right),    \label{2d-e1h4eq1:1}  \\
   \tilde{H} &=& \frac{Q_2^2}{\alpha Q_2^2+\beta}\,\left(P_1^2+P_2^2+\frac{P_3^2}{16 Q_2^2}\right).    \label{2d-e1h4eq1:2}
\eea
The second of these is the $D_2$ kinetic energy with the potential of ``type A'' (parameter $a_3$), in the classification of \cite{03-11}.  Since this corresponds to reduction with respect to $h_4$, which commutes with all our integrals, the latter can be reduced to the 2D space.  We have that $P_1$ is an integral (not just a Killing vector of the metric) and the integrals $F_1$ and $F_2$ take the forms
\be\label{e1h4eq1-F1F2-red}
  \tilde{F}_1={\cal T}_{h_1}^2-4\alpha(Q_1^2+Q_2^2)\tilde{H},\quad \tilde{F}_2= P_1{\cal T}_{h_1}+2\alpha Q_1\tilde{H},
\ee
\end{subequations}
where ${\cal T}_{h_1}$ is a conformal symmetry from the list (\ref{e1h4-conf-h4}).  These integrals satisfy the Poisson relations (\ref{e1h4eq1-pb}) and constraint (\ref{e1h4eq1-constraint}), after the replacement $(e_1,h_4,F_i)\mapsto (P_1,P_3,\tilde F_i)$.

The reduction (\ref{2d-e1h4eq1:1}) possesses a Killing vector, corresponding to the Noether constant $P_3$, but no other quadratic integrals, so is certainly not of Darboux-Koenigs type.

\subsubsection{The Case $\psi(z)=\frac{1}{\alpha z+\beta}$}\label{sec:e1h4eq2-red}

With
\begin{subequations}
\be\label{e1h4eq2-red}
   H=\frac1{\alpha(q_2^2+q_3^2)+\beta}(p_1^2+p_2^2+p_3^2),
\ee
we find that the Hamiltonians (\ref{2d-e1h4:1}) and (\ref{2d-e1h4:2}) give
\bea
\tilde{H} &=& \frac{\text{e}^{-8Q_2}}{16(\alpha \text{e}^{8Q_2}+\beta)}\left(P_2^2+P_3^2+ 16 e^{8Q_2} P_1^2\right),    \label{2d-e1h4eq2:1}  \\
    \tilde{H} &=& \frac1{\alpha Q_2^2+\beta}\left(P_1^2+P_2^2+\frac{P_3^2}{16Q_2^2}\right).    \label{2d-e1h4eq2:2}
\eea
Now the first of these is the $D_3$ kinetic energy with the potential of ``type B'' (parameter $b_3$), in the classification of \cite{03-11}.  Since this corresponds to reduction with respect to $e_1$, which commutes with all our integrals, the latter can be reduced to the 2D space.  The integrals $F_1$ and $F_2$ take the forms
\be\label{e1h4eq2-F1F2-red}
\tilde{F}_1 = -\,\frac{1}{2}{\cal T}_{f_1} {\cal T}_{f_2} +\frac{\alpha}{2} \text{e}^{8Q_2}\sin{(8Q_3)}\tilde{H}, \quad
                    \tilde{F}_2 =  \frac{1}{4}{\cal T}_{f_2}^2 - {\cal T}_{f_1}^2  -\alpha \text{e}^{8Q_2}\cos{(8Q_3)}\tilde{H},
\ee
\end{subequations}
where ${\cal T}_{f_i}$ are conformal symmetries from the list (\ref{e1h4-conf-e1}).  These integrals satisfy the Poisson relations (\ref{e1h4eq2-pb}) and constraint (\ref{e1h4eq2-constraint}), after the replacement $(e_1,h_4,F_i)\mapsto (P_1,P_3,\tilde F_i)$.

The reduction (\ref{2d-e1h4eq2:2}) possesses a Killing vector, corresponding to the Noether constant $P_1$, but no other quadratic integrals, so is certainly not of Darboux-Koenigs type.

\subsubsection{The Case $\psi(z)=\frac{\sqrt{z}}{\alpha \sqrt{z}+\beta}$}\label{sec:e1h4eq3-red}

With
\begin{subequations}
\be\label{e1h4eq3-red}
   H=\frac{\sqrt{q_2^2+q_3^2}}{\alpha\sqrt{q_2^2+q_3^2}+\beta} \left(p_1^2+p_2^2+p_3^2\right),
\ee
we find that the Hamiltonians (\ref{2d-e1h4:1}) and (\ref{2d-e1h4:2}) give
\bea
\tilde{H} &=& \frac{\text{e}^{-4Q_2}}{16(\alpha \text{e}^{4Q_2}+\beta)}\left(P_2^2+P_3^2+16\text{e}^{8Q_2}P_1^2\right),    \label{2d-e1h4eq3:1}  \\
   \tilde{H} &=& \frac{Q_2}{\alpha Q_2+\beta}\left(P_1^2+P_2^2+\frac{P_3^2}{16Q_2^2}\right).    \label{2d-e1h4eq3:2}
\eea
The first of these is the $D_3$ kinetic energy with a potential.  Since this corresponds to reduction with respect to $e_1$, which commutes with all our integrals, the latter can be reduced to the 2D space.  The integrals $F_1$ and $F_2$ take the forms
\be\label{e1h4eq3-F1F2-red}
\tilde{F}_1 = -\,\frac12\,P_3{\cal T}_{f_2}+2\beta\sin{(4Q_3)}\tilde{H},\quad  \tilde{F}_2 = P_3{\cal T}_{f_1}+2\beta\cos{(4Q_3)}\tilde{H},
\ee
\end{subequations}
where ${\cal T}_{f_i}$ are conformal symmetries from the list (\ref{e1h4-conf-e1}).  These integrals satisfy the Poisson relations (\ref{e1h4eq3-pb}) and constraint (\ref{e1h4eq3-constraint}), after the replacement $(e_1,h_4,F_i)\mapsto (P_1,P_3,\tilde F_i)$.

The reduction (\ref{2d-e1h4eq3:2}) possesses a Killing vector, corresponding to the Noether constant $P_1$, but no other quadratic integrals, so is certainly not of Darboux-Koenigs type.

\subsection{Systems with Isometry Algebra $\left<h_1,h_2\right>$}\label{2D-super-reduce-h1h2}

We give a pair of {\em universal} reductions of the general Hamiltonian in this class:
\begin{subequations}
\be\label{3d-h1h2}
H = (q_1^2+q_2^2)\psi\left(\frac{q_3^2}{q_1^2+q_2^2}\right)\, \left(p_1^2+p_2^2+p_3^2\right).
\ee
We then give the reductions of the four cases presented in Section \ref{2D-super-h1h2}.

Since the isometries {\em commute}, we can {\em simultaneously} reduce them to $(h_1,h_2)=(P_1,P_3)$, giving us the form of $Q_1, Q_3$, up to arbitrary functions of the common invariant $z = \frac{q_3^2}{q_1^2+q_2^2}$.  We solve the equations
\bea
 \{Q_1,h_1\}=1,\;\;\; \{Q_1,h_2\}=0 &\Rightarrow & Q_1 = -\frac{1}{4}\, \log{\left(q_1^2+q_2^2\right)}+w_1(z),  \nn\\
\{Q_2,h_1\}=0,\;\;\; \{Q_2,h_2\}=0 &\Rightarrow & Q_2 = w_2(z),  \nn\\
\{Q_3,h_1\}=0,\;\;\; \{Q_3,h_2\}=1 &\Rightarrow & Q_3 = -\frac{1}{2}\, \arctan\left(\frac{q_1}{q_2}\right)+w_3(z),  \nn
\eea
to obtain:
\be\label{h1h2-genS}
S = \left(-\frac{1}{4}\, \log{\left(q_1^2+q_2^2\right)}+w_1(z)\right)\,P_1+w_2(z)\,P_2+\left(-\frac{1}{2}\, \arctan\left(\frac{q_1}{q_2}\right)+w_3(z)\right)\,P_3.
\ee
We first require that the coefficients of $P_iP_j$ are zero, giving
$$
(1+4 (1+z)w_1'(z))w_2'(z)=w_2'(z)w_3'(z)=(1+4 (1+z)w_1'(z))w_3'(z)=0 ,
$$
with solution $w_1(z)  = -\frac{1}{4}\, \log (1+z),\;  w_3(z)  =  0$, since $w_2'(z)$ {\em cannot} be zero.  This leaves only $w_2(z)$ arbitrary in
\be\label{h1h2-Sw2}
S=-\frac{1}{4}\,\log{\left(q_1^2+q_2^2+q_3^2\right)}\;P_1+w_2(z)\;P_2-\frac{1}{2}\, \arctan\left(\frac{q_1}{q_2}\right)\;P_3,
\ee
\end{subequations}
with the variable $Q_2$ being chosen by equating coefficients of $P_1^2,P_2^2$ or of $P_2^2,P_3^2$, giving an ODE for $w_2(z)$ in each case, leading to the two transformations.

\subsubsection*{Reduction with respect to $h_1$}

Here we have the generating function
\begin{subequations}
\be\label{TRh1h2:h1}
S = -\frac{1}{4}\,\log\left(q_1^2+q_2^2+q_3^2\right)\,P_1+\frac12\, \log\left(\frac{q_3+\sqrt{q_1^2+q_2^2+q_3^2}}{\sqrt{q_1^2+q_2^2}}\right)\,P_2-\frac12\arctan\left(\frac{q_1}{q_2}\right)\,P_3,
\ee
giving the Hamiltonian
\be  \label{2d-h1h2:h1}
 H = \frac{1}{4}\,  \psi\left(\sinh^2 2 Q_2\right) \left(P_2^2+P_3^2+ \mbox{sech}^2\,(2 Q_2)\,  P_1^2\right).
\ee
This case corresponds to a conformally flat metric in the $2-3$ space, defined by $P_1 = \mbox{const.}$, with $\mbox{sech}^2\,(2 Q_2)\,  P_1^2$ being an additional potential term.  Since the conformal factor is a function of only $Q_2$, momentum $P_3$ corresponds to a Killing vector (in 2D), as well as a first integral of the entire Hamiltonian.

Again, we build the 6 dimensional conformal algebra for this 2D metric (kinetic energy).  First note that $h_2,\, h_3,\, h_4,\, e_1f_1+\frac{1}{4} h_1^2,\, e_2 f_2,\,  e_3 f_3$ {\em commute} with $h_1$.
Writing these in terms of $Q_i,P_i$ and discarding the $P_1^2$ components in the quadratic ones, we can derive the following 6 conformal elements:
\bea
&&  {\cal T}_{e_1} = \frac{1}{2} e^{2Q_2} (P_2 \sin 2 Q_3-P_3 \cos 2Q_3),\quad {\cal T}_{h_1} = P_2,\quad {\cal T}_{f_1}= -\frac{1}{2} e^{-2Q_2} (P_2 \sin 2 Q_3+P_3 \cos 2Q_3), \nn  \\[-1mm]
&&       \label{h1h2-confh1}   \\[-1mm]
&&  {\cal T}_{e_2} = \frac{1}{2} e^{2Q_2} (P_2 \cos 2 Q_3+P_3 \sin 2Q_3),\quad {\cal T}_{h_2} = -P_3,\quad {\cal T}_{f_2}=  e^{-2Q_2} (P_3 \sin 2Q_3- P_2 \cos 2 Q_3),  \nn
\eea
\end{subequations}
which satisfy the relations of $\mathfrak{g}_1 + \mathfrak{g}_2$ in Table \ref{Tab:g1234}, together with the algebraic constraints (\ref{e1h4-conf-constraints}).

\subsubsection*{Reduction with respect to $h_2$}

Here we have the generating function
\begin{subequations}
\be\label{TRh1h2:h2}
S =-\frac{1}{4}\,\log{\left(q_1^2+q_2^2+q_3^2\right)}\,P_1+\frac12\arctan\left(\frac{q_3}{\sqrt{q_1^2+q_2^2}}\right)\,P_2-\frac12\arctan\left(\frac{q_1}{q_2}\right)\,P_3,
\ee
giving the Hamiltonian
\be   \label{2d-h1h2:h2}
 H = \frac{1}{4}\, \cos^2(2 Q_2)\,\psi\left(\tan^2 2 Q_2\right) \left(P_1^2+P_2^2+\sec^2(2 Q_2) P_3^2\right).
\ee
This case corresponds to a conformally flat metric in the $1-2$ space, defined by $P_3 = \mbox{const.}$, with $\sec^2(2 Q_2) P_3^2$ being an additional potential term.  Since the conformal factor is a function of only $Q_2$, the momentum $P_1$ corresponds to a Killing vector (in 2D), and, again, is a first integral of the entire Hamiltonian.

Again, we build the 6 dimensional conformal algebra for this 2D metric (kinetic energy).  First note that $e_3,\, h_1,\, f_3,\, e_1^2+e_2^2,\, h_3^2+\frac{1}{4} h_4^2,\,  f_2^2+4 f_1^2$ {\em commute} with $h_2$.
Writing these in terms of $Q_i,P_i$ and discarding the $P_3^2$ components in the quadratic ones, we can derive the following 6 conformal elements:
\bea
&&  {\cal T}_{e_1} = \frac{1}{2} e^{2Q_1} (P_2 \cos 2 Q_2-P_1 \sin 2Q_2),\quad {\cal T}_{h_1} = P_1,\quad {\cal T}_{f_1}=\frac{1}{2} e^{-2Q_1} (P_2 \cos 2 Q_2+P_1 \sin 2Q_2), \nn  \\[-1mm]
&&       \label{h1h2-confh2}   \\[-1mm]
&&  {\cal T}_{e_2} = \frac{1}{2} e^{2Q_1} (P_2 \sin 2 Q_2+P_1 \cos 2Q_1),\quad {\cal T}_{h_2} = P_2,\quad {\cal T}_{f_2}=  e^{-2Q_1} (P_2 \sin 2 Q_2-P_1 \cos 2Q_2),  \nn
\eea
\end{subequations}
which satisfy the relations of $\mathfrak{g}_1 + \mathfrak{g}_2$ in Table \ref{Tab:g1234}, together with the algebraic constraints (\ref{e1h4-conf-constraints}).

\subsubsection{The Case $\psi(z)=\frac{(1+z) z}{\alpha +\beta z}$}\label{sec:h1h2eq1-red}

With
\begin{subequations}
\be\label{h1h2eq1-red}
   H=\frac{\left(q_1^2+q_2^2+q_3^2\right)q_3^2}{\alpha\left(q_1^2+q_2^2\right)+\beta q_3^2}\left(p_1^2+p_2^2+p_3^2\right),
\ee
we find that the Hamiltonians (\ref{2d-h1h2:h1}) and (\ref{2d-h1h2:h2}) give
\bea
\tilde{H} &=& \frac{\sinh^2{(4Q_2)}}{8(\beta\cosh{(4Q_2)}+2\alpha-\beta)}\left(P_2^2+P_3^2+  \mbox{sech}^2\,2 Q_2\, P_1^2\right),    \label{2d-h1h2eq1:2}   \\
  \tilde{H} &=& \frac{\sin^2{(2Q_2)}}{4(\alpha\cos^2{(2Q_2)}+\beta\sin^2{(2Q_2)})}  \left(P_1^2+P_2^2+\sec^2{(2Q_2)} P_3^2\right).    \label{2d-h1h2eq1:1}
\eea
The first of these is the $D_4$ kinetic energy (but with {\em hyperbolic} functions in place of trigonometric ones) with a potential.  Since this corresponds to reduction with respect to $h_1$, which commutes with all our integrals, the latter can be reduced to the 2D space.  The integrals $F_1$ and $F_2$ take the forms
\be\label{h1h2eq1-F1F2-red}
\tilde{F}_1 = \tilde{h}_3\tilde{h}_4-4\alpha\frac{\sin{(4Q_3)}}{\sinh^2{(2Q_2)}}\tilde{H},\quad    \tilde{F}_2 = 4\tilde{h}_3^2-\tilde{h}_4^2+16\alpha\frac{\cos{(4Q_3)}}{\sinh^2{(2Q_2)}}\tilde{H},
\ee
\end{subequations}
where $\tilde h_3 = {\cal T}_{e_1}-{\cal T}_{f_1},\; \tilde h_4 = 2{\cal T}_{e_2}-{\cal T}_{f_2}$, with ${\cal T}_{e_i},\, {\cal T}_{f_i}$ being conformal symmetries from the list (\ref{h1h2-confh1}).
These integrals satisfy the Poisson relations (\ref{h1h2eq1-pb}) and constraint (\ref{h1h2eq1-constraint}), after the replacement $(h_1,h_2,F_i)\mapsto (P_1,P_3,\tilde F_i)$.

The reduction (\ref{2d-h1h2eq1:1}) possesses a Killing vector, corresponding to the Noether constant $P_1$, but no other quadratic integrals, so is not of Darboux-Koenigs type.

\subsubsection{The Case $\psi(z)=\frac{\sqrt{1+z}}{\alpha \sqrt{1+z} +\beta \sqrt{z}}$}\label{sec:h1h2eq2-red}

With
\begin{subequations}
\be\label{h1h2eq2-red}
   H=\frac{\sqrt{q_1^2+q_2^2+q_3^2}\, \left(q_1^2+q_2^2\right)}{\alpha\sqrt{q_1^2+q_2^2+q_3^2}+\beta q_3}\, \left(p_1^2+p_2^2+p_3^2\right),
\ee
we find that the Hamiltonians (\ref{2d-h1h2:h1}) and (\ref{2d-h1h2:h2}) give
\bea
 \tilde{H} &=& \frac{\cosh{(2Q_2)}}{4(\alpha\cosh{(2Q_2)}+\beta\sinh{(2Q_2)})}    \left(P_2^2+P_3^2+ \mbox{sech}^2\,2 Q_2\, P_1^2\right),    \label{2d-h1h2eq2:2}\\
  \tilde{H} &=& \frac{\cos^2{(2Q_2)}}{4(\alpha+\beta\sin{(2Q_2)})} \left(P_1^2+P_2^2+\sec^2{(2Q_2)} P_3^2\right).    \label{2d-h1h2eq2:1}
\eea
The second of these is the $D_4$ kinetic energy $\left(\mbox{after a shift}\; Q_2\rightarrow Q_2-\frac{\pi}{4}\right)$ with a potential of ``type A" (parameter $a_2$) in the classification of \cite{03-11}.  Since this corresponds to reduction with respect to $h_2$, which commutes with all our integrals, the latter can be reduced to the 2D space.  The integrals $F_1$ and $F_2$ take the forms
\be\label{h1h2eq2-F1F2-red}
\tilde{F}_1 = P_1{\cal T}_{e_1}-\beta\text{e}^{2Q_1}\tilde{H},\quad    \tilde{F}_2 = 2P_1{\cal T}_{f_1} +2\beta\text{e}^{-2Q_1}\tilde{H},
\ee
\end{subequations}
where ${\cal T}_{e_1},\, {\cal T}_{f_1}$ are conformal symmetries from the list (\ref{h1h2-confh2}).  These integrals satisfy the Poisson relations (\ref{h1h2eq2-pb}) and constraint (\ref{h1h2eq2-constraint}), after the replacement $(h_1,h_2,F_i)\mapsto (P_1,P_3,\tilde F_i)$.

The reduction (\ref{2d-h1h2eq2:2}) possesses a Killing vector, corresponding to the Noether constant $P_3$, but no other quadratic integrals, so is not of Darboux-Koenigs type.

\subsubsection{The Case $\psi(z)=\frac{1+z}{\alpha  +\beta \sqrt{z}}$}\label{sec:h1h2eq3-red}

With
\begin{subequations}
\be\label{h1h2eq3-red}
   H=\frac{\sqrt{q_1^2+q_2^2}\, \left(q_1^2+q_2^2+q_3^2\right)}{\alpha\sqrt{q_1^2+q_2^2}+\beta q_3}\, \left(p_1^2+p_2^2+p_3^2\right),
\ee
we find that the Hamiltonians (\ref{2d-h1h2:h1}) and (\ref{2d-h1h2:h2}) give
\bea
\tilde{H} &=& \frac{\cosh^2{(2Q_2)}}{4(\alpha+\beta\sinh{(2Q_2)})}\left(P_2^2+P_3^2+\mbox{sech}^2\,2 Q_2\, P_1^2\right),    \label{2d-h1h2eq3:2}  \\
  \tilde{H} &=& \frac{\cos{(2Q_2)}}{4(\alpha\cos{(2Q_2)}+\beta\sin{(2Q_2)})}  \left(P_1^2+P_2^2+\sec^2{(2Q_2)} P_3^2\right).    \label{2d-h1h2eq3:1}
\eea
The first of these is the $D_4$ kinetic energy (but with {\em hyperbolic} functions in place of trigonometric) with a potential.  Since this corresponds to reduction with respect to $h_1$, which commutes with all our integrals, the latter can be reduced to the 2D space.  The integrals $F_1$ and $F_2$ take the forms
\be \label{h1h2eq3-F1F2-red}
\tilde{F}_1 = P_3\left({\cal T}_{f_1}-{\cal T}_{e_1}\right)-2\beta\cos{(2Q_3)}\tilde{H},\quad   \tilde{F}_2 = P_3\left({\cal T}_{f_2}-2{\cal T}_{e_2}\right)+4\beta\sin{(2Q_3)}\tilde{H},
\ee
\end{subequations}
where ${\cal T}_{e_i},\, {\cal T}_{f_i}$ are conformal symmetries from the list (\ref{h1h2-confh1}).  These integrals satisfy the Poisson relations (\ref{h1h2eq3-pb}) and constraint (\ref{h1h2eq3-constraint}), after the replacement $(h_1,h_2,F_i)\mapsto (P_1,P_3,\tilde F_i)$.

The reduction (\ref{2d-h1h2eq3:1}) possesses a Killing vector, corresponding to the Noether constant $P_1$, but no other quadratic integrals, so is not of Darboux-Koenigs type.

\subsubsection{The Case $\psi(z)=\frac{z}{\alpha +\beta z}$}\label{sec:h1h2eq4-red}

With
\begin{subequations}
\be\label{h1h2eq4-red}
   H=\frac{\left(q_1^2+q_2^2\right)q_3^2}{\alpha\left(q_1^2+q_2^2\right)+\beta q_3^2}\, \left(p_1^2+p_2^2+p_3^2\right),
\ee
we find that the Hamiltonians (\ref{2d-h1h2:h1}) and (\ref{2d-h1h2:h2}) give
\bea
\tilde{H} &=& \frac{\sinh^2{(2Q_2)}}{4(\alpha+\beta\sinh^2{(2Q_2)})}  \left(P_2^2+P_3^2+\mbox{sech}^2\,2 Q_2\, P_1^2\right),    \label{2d-h1h2eq4:2}  \\
  \tilde{H} &=& \frac{\sin^2{(4Q_2)}}{8((\alpha-\beta)\cos{(4Q_2)}+\alpha+\beta)}   \left(P_1^2+P_2^2+\sec^2{(2Q_2)} P_3^2\right).    \label{2d-h1h2eq4:1}
\eea
The second of these is the $D_4$ kinetic energy with a potential.  Since this corresponds to reduction with respect to $h_2$, which commutes with all our integrals, the latter can be reduced to the 2D space.  The integrals $F_1$ and $F_2$ take the forms
\be \label{h1h2eq4-F1F2-red}
\tilde{F}_1 = {\cal T}_{e_1}^2-\frac{\alpha\text{e}^{4Q_1}}{\sin^2{(2Q_2)}}\tilde{H},\quad   \tilde{F}_2 = 4{\cal T}_{f_1}^2-\frac{4\alpha\text{e}^{-4Q_1}}{\sin^2{(2Q_2)}}\tilde{H},
\ee
\end{subequations}
where ${\cal T}_{e_1},\, {\cal T}_{f_1}$ are conformal symmetries from the list (\ref{h1h2-confh2}).  These integrals satisfy the Poisson relations (\ref{h1h2eq4-pb}) and constraint (\ref{h1h2eq4-constraint}), after the replacement $(h_1,h_2,F_i)\mapsto (P_1,P_3,\tilde F_i)$.

The reduction (\ref{2d-h1h2eq4:2}) possesses a Killing vector, corresponding to the Noether constant $P_3$, but no other quadratic integrals, so is not of Darboux-Koenigs type.

\section{Reduction to 2 Dimensional Systems: 3D Isometry Algebras}\label{3D-super-reduce}

In this section we use the approach described in Section \ref{sec:reductions} to reduce our systems of Section \ref{3D-super} to 2 dimensional spaces.  This time the isometries do not form a commutative algebra, so we cannot {\em simultaneously} adapt coordinates to more than one isometry.  Nevertheless, we can still reduce the systems in a ``universal'' way, transforming each Hamiltonian into  a pair of ``conformal forms'' in the remaining 2 dimensions.  In fact, the conformal factor is the same for both reductions, but the added ``potential'' is different.

Any integrals which {\em commute} with this isometry will reduce to give an integral in the two dimensional space.  Hence, our superintegrable cases reduce to superintegrable systems in 2D, with one Killing vector and two second order Killing tensors, so reduce to Darboux-Koenigs systems.

\subsection{Systems with Isometry Algebra $\left<e_1,h_1,f_1\right>$}\label{2D-super-reduce-e1h1f1}

We give a pair of {\em universal} reductions of the general Hamiltonian in this class:
\be\label{3d-e1h1f1}
H = q_2^2 \psi\left(\frac{q_3}{q_2}\right) \left(p_1^2+p_2^2+p_3^2\right).
\ee
We then give the reductions of the two cases presented in Section \ref{3D-super-e1h1f1}.

We give two transformations, corresponding to $e_1\rightarrow P_1$ and $h_1\rightarrow P_3$, respectively.  The transformation using $f_1$ can be obtained from that involving $e_1$ by applying the involution $\iota_{ef}$, which preserves the general Hamiltonian.

In each case, we choose $Q_2$ to be a function of $z_2=\frac{q_3}{q_2}$, which is the common invariant of $e_1$ and $h_1$ and, as a consequence, the variable which appears in the arbitrary function $\psi$, of the Hamiltonian. In fact, it turns out that $Q_2$ is exactly the {\em same} function of $z_2$ in each case, so that the conformal factor of the reduced metric is the same in each case.

\subsubsection{Reduction using the Isometry $e_1\mapsto P_1$}

In view of $\{q_1,e_1\}=1$ and $\{q_2,e_1\}=\{q_3,e_1\}=0$, we consider the generating function
\begin{subequations}
\be\label{e1h1f1-S1e1}
  S=(q_1+w_1(q_2,q_3))\;P_1+w_2(z_2)\;P_2+w_3(q_2,q_3)\;P_3, \quad\mbox{where}\;\;\; z_2 = \frac{q_3}{q_2}.
\ee
We first require that the coefficients of $P_iP_j$ are zero, giving
$$
w_2'(z_2)\left(q_2\pa_{q_3}w_1-q_3\pa_{q_2}w_1\right) = w_2'(z_2)\left(q_2\pa_{q_3}w_3-q_3\pa_{q_2}w_3\right) = \pa_{q_2}w_1\pa_{q_2}w_3+\pa_{q_3}w_1\pa_{q_3}w_3 = 0.
$$
The first two give $w_i=w_i\left(q_2^2+q_3^2\right),\; i=1,3$, since $w_2'(z_2)$ {\em cannot} be zero.  The third then gives $w_1'w_3'=0$.
Since $w_3'$ \underline{cannot} be zero, we have $w_1=0$, so the canonical transformation (\ref{e1h1f1-S1e1}) now takes the form
\be\label{e1h1f1-S2e1}
  S= q_1\;P_1+w_2(z_2)\;P_2+w_3(z_1)\;P_3, \quad\mbox{where}\;\;\; z_1= q_2^2+q_3^2 , \;\; z_2 = \frac{q_3}{q_2}.
\ee
At this stage, the transformed Hamiltonian is diagonal:
\be\label{e1h1f1-Ht1e1}
  \tilde{H}=(1+z_2^2)\psi(z_2)w_2'^2(z_2)\;P_2^2+\frac{4\psi(z_2)}{1+z_2^2}z_1^2 w_3'^2(z_1)\;P_3^2+\frac{z_1\psi(z_2)}{1+z_2^2}\;P_1^2.
\ee
We must choose $w_2$ and $w_3$ so that (\ref{e1h1f1-Ht1e1}) is ``conformal'' in the $P_2-P_3$ components, which requires
$$
   4 z_1^2 w_3'^2(z_1)=1,\qquad(1+z_2^2)w_2'^2(z_2)=\frac1{1+z_2^2},
$$
giving
$$
  w_2=\arctan{z_2}=\arctan\left(\frac{q_3}{q_2}\right),\quad w_3=\frac12\log{z_1}=\frac12\log\left(q_2^2+q_3^2\right).
$$
This gives us the final form of the canonical transformation
\be\label{e1h1f1-S3e1}
  S=q_1\;P_1+\arctan \left(\frac{q_3}{q_2}\right)\;P_2+\frac12\log \left(q_2^2+q_3^2\right)\;P_3,
\ee
and the Hamiltonian
\be\label{e1h1f1-Ht2e1}
  \tilde{H}=\cos^2{Q_2}\;\psi(\tan{Q_2})\left(P_2^2+P_3^2+\text{e}^{2Q_3}P_1^2\right).
\ee
This corresponds to a conformally flat metric in the $2-3$ space, defined by $P_1=\mbox{const}$, with $\text{e}^{2Q_3}P_1^2$ corresponding to a potential term.  Since the conformal factor is a function of only $Q_2$, the momentum $P_3$ corresponds to a Killing vector (in 2D), but not a first integral of the entire Hamiltonian.  However, the Casimir function of the original isometry algebra \underline{does} reduce to a first integral of $\tilde H$:
\be\label{e1h1f1-case1}
{\cal C} = e_1f_1+\frac14h_1^2 = P_3^2+\text{e}^{2Q_3}P_1^2.
\ee
Thus, for \underline{arbitrary} $\psi$, the Hamiltonian $\tilde H$ is Liouville integrable and, indeed, separable.  The form of the potential and of the Casimir are \underline{independent} of the form of $\psi$, so are \underline{universal} properties of this reduction.

Again, we build the 6 dimensional conformal algebra for this 2D metric (kinetic energy).  First note that $e_2,\, e_3,\, h_4,\, e_1f_1+\frac{1}{4} h_1^2,\, e_2 f_2-\frac{1}{2} h_2^2,\,  e_3 f_3-\frac{1}{2} h_3^2$ {\em commute} with $e_1$.
Writing these in terms of $Q_i,P_i$ and discarding the $P_1^2$ components in the quadratic ones, we can derive the following 6 conformal elements:
\bea
&&  {\cal T}_{e_1} = e^{Q_3} (P_2 \sin Q_2+P_3 \cos Q_2),\quad {\cal T}_{h_1} = 2 P_3,\quad {\cal T}_{f_1}=e^{-Q_3} (P_2 \sin Q_2-P_3 \cos Q_2),  \nn  \\[-1mm]
&&                             \label{e1h1f1-conf-e1}   \\[-1mm]
&&  {\cal T}_{e_2} = e^{Q_3} (P_2 \cos Q_2-P_3 \sin Q_2),\quad {\cal T}_{h_2} = -2 P_2,\quad {\cal T}_{f_2}= 2e^{-Q_3} (P_2 \cos Q_2+P_3 \sin Q_2), \nn
\eea
\end{subequations}
which satisfy the relations of $\mathfrak{g}_1 + \mathfrak{g}_2$ in Table \ref{Tab:g1234}, together with the algebraic constraints (\ref{e1h4-conf-constraints}).

\subsubsection{Reduction using the Isometry $h_1\mapsto P_3$}

In view of $\left\{-\frac12\log{q_1},h_1\right\}=1$ and $\left\{\frac{q_2}{q_1},h_1\right\}=\left\{\frac{q_3}{q_1},h_1\right\}=0$, we consider the generating function
\begin{subequations}
\be\label{e1h1f1-S1h1}
  S=w_1(z_1,z_3)\;P_1+w_2(z_2)\;P_2+\left(-\frac12\log{q_1}+w_3(z_1,z_3)\right)\;P_3,
\ee
where $z_1=\frac{q_2}{q_1},\;\; z_2 = \frac{q_3}{q_2},\;\; z_3=\frac{q_3}{q_1}$.  We first require that the coefficients of $P_iP_j$ are zero, giving
\bea
&&  w_2'(z_2)\left(z_1\pa_{z_3}w_1-z_3\pa_{z_1}w_1\right) = w_2'(z_2)\left(z_1\pa_{z_3}w_3-z_3\pa_{z_1}w_3\right)  = 0,  \nn\\[2mm]
&&   \left((2(1+z_1^2)\pa_{z_1}w_3+2z_1z_3\pa_{z_3}w_3+z_1)\pa_{z_1}w_1+  (2(1+z_3^2)\pa_{z_3}w_3+2z_1z_3\pa_{z_1}w_3+z_3)\pa_{z_3}w_1\right)=0.  \nn
\eea
The first two give
$$
w_i=w_i(z_1^2+z_3^2)=w_i\left(\frac{q_2^2+q_3^2}{q_1^2}\right),\qquad i=1,3,
$$
since $w_2'(z_2)$ {\em cannot} be zero.  The third then gives
$$
\left(4(1+z_4)w_3'(z_4)+1\right)w_1'(z_4)=0,\quad\mbox{where}\;\;\; z_4=\frac{q_2^2+q_3^2}{q_1^2}.
$$
Since $w_1'$ \underline{cannot} be zero, we have
$$
w_3=-\frac14\log{(1+z_4)}=-\frac14\log\left(\frac{q_1^2+q_2^2+q_3^2}{q_1^2}\right),
$$
so the canonical transformation (\ref{e1h1f1-S1h1}) now takes the form
\be\label{e1h1f1-S2h1}
   S=w_1\left(\frac{q_2^2+q_3^2}{q_1^2}\right)\;P_1+w_2\left(\frac{q_3}{q_2}\right)\;P_2-\frac14\log\left(q_1^2+q_2^2+q_3^2\right)\;P_3.
\ee
At this stage, the transformed Hamiltonian is diagonal:
\be\label{e1h1f1-Ht1h1}
 \tilde{H}=\frac{4 \psi(z_2)}{1+z_2^2}(1+z_4)z_4^2w_1'^2(z_4)\;P_1^2+(1+z_2^2)\psi(z_2) w_2'^2(z_2)\;P_2^2+\frac{z_4 \psi(z_2)}{4(1+z_2^2)(1+z_4)}\;P_3^2.
\ee
We must choose $w_1$ and $w_2$ so that this Hamiltonian is ``conformal'' in the $P_1-P_2$ components, which requires
$$
   (1+z_4)z_4^2w_1'^2(z_4)=\frac14,\quad  (1+z_2^2)w_2'^2(z_2)=\frac1{1+z_2^2},
$$
giving
$$
  w_1=\frac12 \log\left(\frac{\sqrt{1+z_4}+1}{\sqrt{1+z_4}-1}\right)=\frac{1}{2}\, \log\left(\frac{\sqrt{q_1^2+q_2^2+q_3^2}+q_1}{\sqrt{q_1^2+q_2^2+q_3^2}-q_1}\right),\quad w_2=\arctan{z_2}=\arctan\left(\frac{q_3}{q_2}\right).
$$
This gives us the final form of the canonical transformation
\be\label{e1h1f1-S3h1}
  S=\frac{1}{2}\, \log\left(\frac{\sqrt{q_1^2+q_2^2+q_3^2}+q_1}{\sqrt{q_1^2+q_2^2+q_3^2}-q_1}\right)\;P_1+\arctan\left(\frac{q_3}{q_2}\right)\;P_2-\frac14\log\left(q_1^2+q_2^2+q_3^2\right)\;P_3,
\ee
and the Hamiltonian
\be\label{e1h1f1-Ht2h1}
  \tilde{H}=\cos^2{Q_2}\;\psi\left(\tan{Q_2}\right)\left(P_1^2+P_2^2+\frac1{4\cosh^2Q_1}P_3^2\right).
\ee
This corresponds to a conformally flat metric in the $1-2$ space, defined by $P_3=\mbox{const}$, with $\frac1{4\cosh^2Q_1}P_3^2$ corresponding to a potential term.  Since the conformal factor is a function of only $Q_2$, the momentum $P_1$ corresponds to a Killing vector (in 2D), but not a first integral of the entire Hamiltonian.  However, the Casimir function of the original isometry algebra \underline{does} reduce to a first integral of $\tilde H$:
\be\label{e1h1f1-cash1}
e_1f_1+\frac14h_1^2 = P_1^2+\frac1{4\cosh^2Q_1}P_3^2.
\ee
Thus, for \underline{arbitrary} $\psi$, the Hamiltonian $\tilde H$ is Liouville integrable and, indeed, separable.  The form of the potential and of the Casimir are \underline{independent} of the form of $\psi$, so are \underline{universal} properties of this reduction.

Again, we build the 6 dimensional conformal algebra for this 2D metric (kinetic energy).  First note that $h_2,\, h_3,\, h_4,\, e_1f_1+\frac{1}{4} h_1^2,\, e_2 f_2,\, e_3 f_3$ {\em commute} with $h_1$.
Writing these in terms of $Q_i,P_i$ and discarding the $P_3^2$ components in the quadratic ones, we can derive the following 6 conformal elements:
\bea
&&  {\cal T}_{e_1} = e^{Q_1} (P_2 \sin Q_2+P_1 \cos Q_2),\quad {\cal T}_{h_1} = 2 P_1,\quad {\cal T}_{f_1}=e^{-Q_1} (P_2 \sin Q_2-P_1 \cos Q_2), \nn  \\[-1mm]
&&                                                            \label{e1h1f1-conf-h1}    \\[-1mm]
&&  {\cal T}_{e_2} = e^{Q_1} (P_2 \cos Q_2-P_1 \sin Q_2),\quad {\cal T}_{h_2} = -2 P_2,\quad {\cal T}_{f_2}= 2e^{-Q_1} (P_2 \cos Q_2+P_1 \sin Q_2),  \nn
\eea
\end{subequations}
which satisfy the relations of $\mathfrak{g}_1 + \mathfrak{g}_2$ in Table \ref{Tab:g1234}, together with the algebraic constraints (\ref{e1h4-conf-constraints}).

\subsubsection{Specific Cases Listed in Section \ref{3D-super-e1h1f1}}

In Section \ref{3D-super-e1h1f1} we gave two specific choices of the function $\psi$, which allowed the addition of quadratic integrals.  A detailed analysis of the reduction of these was given in \cite{f19-2}, so we just include a brief discussion here to fit these into the above framework.  To obtain a superintegrable system in the 2 dimensional context, we need two first integrals which commute with the reducing isometry.  One of these can be the Casimir of the isometry algebra, the other coming from the algebra of functions $F_i$.

\subsubsection*{The Case $\psi(z)=\frac{z^2}{\alpha  +\beta z^2}$}

When we reduce equation (\ref{e1h1f1eq1}), using $e_1$, the Hamiltonian (\ref{e1h1f1-Ht2e1}) and Casimir (\ref{e1h1f1-case1}) take the form
\begin{subequations}
\bea
\tilde H &=& \frac{\sin^2(2 Q_2)}{2(\alpha+\beta)+2(\alpha-\beta) \cos(2Q_2)} \, \left(P_2^2+P_3^2+e^{2 Q_3} P_1^2\right),  \label{e1h1f1eq1DK1}  \\
 J_1 &=&  P_3^2+\text{e}^{2Q_3}P_1^2.  \label{e1h1f1eq1DK1J1}
\eea
This is the $D_4$ kinetic energy, with potential (equivalent to the ``$a_1$'' part of Case A in \cite{03-11}).

It is easily seen that $F_3 = e_2^2 -\frac{\beta}{q_2^2}\, H$ is a first integral, satisfying $\{H,F_3\}=\{e_1,F_3\}=0$ (see Table 6 of \cite{f19-2} for the full Poisson algebra).  This gives another (independent) quadratic integral for the 2 dimensional system:
\be\label{e1h1f1eq1DK1J2}
J_2 = {\cal T}_{f_1}^2-\beta e^{-2 Q_3} \sec^2 Q_2\, \tilde H,
\ee
where ${\cal T}_{f_1}$ is a conformal symmetry in the 2 dimensional space, listed in (\ref{e1h1f1-conf-e1}).
\end{subequations}
These integrals, together with the cubic $J_3=\{J_1,J_2\}$, satisfy the relations
\bea
&&  \{J_1,J_3\} = 16 J_1J_2+8 P_1^2 \left(J_1+(\beta-\alpha)  \tilde H\right),\quad  \{J_2,J_3\} = -8J_2 \left(J_2 +P_1^2\right),  \nn\\
&&  J_3^2 = 16 \left(J_1 J_2(J_2+P_1^2)+P_1^2 \tilde H\left((\beta-\alpha) J_2+\beta P_1^2\right)\right).   \nn
\eea

\medskip
When we reduce equation (\ref{e1h1f1eq1}), using $h_1$, the Hamiltonian (\ref{e1h1f1-Ht2h1}) and Casimir (\ref{e1h1f1-cash1}) take the form
\begin{subequations}
\bea
\tilde H &=& \frac{\sin^2(2 Q_2)}{2(\alpha+\beta)+2(\alpha-\beta) \cos(2Q_2)} \, \left(P_1^2+P_2^2+\frac{1}{4}\, \mbox{sech}^2\, Q_1 P_3^2\right),  \label{e1h1f1eq1DK2}  \\
 J_1 &=&  P_1^2+\frac{1}{4}\, \mbox{sech}^2\, Q_1 P_3^2.  \label{e1h1f1eq1DK2J1}
\eea
This is the $D_4$ kinetic energy, with potential (equivalent to the ``$b_3$'' part of Case B in \cite{03-11}).

It is easily seen that the first integral $F_1$ (see (\ref{e1h1f1eq1-F1})) satisfies $\{h_1,F_1\}=0$ (see Table 6 of \cite{f19-2} for the full Poisson algebra).  This gives another (independent) quadratic integral for the 2 dimensional system:
\be\label{e1h1f1eq1DK2J2}
J_2 = \frac{1}{2} ({\cal T}_{e_1}+{\cal T}_{f_1})^2+\beta (\cos 2 Q_2-\cosh 2 Q_1) \sec^2 Q_2\, \tilde H -\frac{1}{2}\, P_3^2 \cos^2 Q_2\, \mbox{sech}^2\, Q_1 ,
\ee
where ${\cal T}_{e_1}$ and ${\cal T}_{f_1}$ are conformal symmetries in the 2 dimensional space, listed in (\ref{e1h1f1-conf-h1}).
\end{subequations}
These integrals, together with the cubic $J_3=\{J_1,J_2\}$, satisfy the relations
\bea
  \{J_1,J_3\} &=& 4 \left(J_1\left(4(J_1+J_2)+ P_3^2\right)-\tilde H \left(4(\alpha+\beta)J_1+(\alpha-\beta)P_3^2  \right)\right),\nn\\
\{J_2,J_3\} &=& -4(4J_1+J_2)(2J_2+P_3^2)+8\tilde H \left(2(\alpha+\beta)J_2 +(2\alpha+\beta)P_3^2\right)-32\alpha\beta \tilde H^2,  \nn\\
  J_3^2 &=& 8J_1(2J_1+J_2)(2J_2+P_3^2)-4\tilde H \left(2\alpha (4 J_1J_2+(J_2+4J_1)P_3^2)\right.  \nn\\
    &&  \hspace{2cm}  \left. +\beta(4J_1-P_3^2)(2J_2+P_3^2)\right) +16 \alpha \tilde H^2(4\beta J_1+(\alpha-\beta)P_3^2).   \nn
\eea

\subsubsection*{The Case $\psi(z)=\frac{\sqrt{1+z^2}}{\alpha \sqrt{1+z^2}  +\beta z}$}

The details of this case can be found in  \cite{f19-2} (Sections 5.1 and 9.1), but see Remark \ref{e1h1f1eq2-rem}.

\medskip
When we reduce equation (\ref{e1h1f1eq2}), using $e_1$, the Hamiltonian (\ref{e1h1f1-Ht2e1}) and Casimir (\ref{e1h1f1-case1}) take the form
\begin{subequations}
\bea
\tilde H &=& \frac{\cos^2(Q_2)}{\alpha+\beta \sin Q_2} \, \left(P_2^2+P_3^2+e^{2 Q_3} P_1^2\right),  \label{e1h1f1eq2DK1}  \\
 J_1 &=&  P_3^2+\text{e}^{2Q_3}P_1^2.  \label{e1h1f1eq2DK1J1}
\eea
This is the $D_4$ kinetic energy, with potential (equivalent to the ``$a_1$'' part of Case A in \cite{03-11}).

It is easily seen that $F_1$ (see (\ref{e1h1f1eq2-F1})) satisfies $\{e_1,F_1\}=0$ (see Table 8 of \cite{f19-2} for the full Poisson algebra).  This gives another (independent) quadratic integral for the 2 dimensional system:
\be\label{e1h1f1eq2DK1J2}
J_2 = 2 P_2 {\cal T}_{f_1}-\frac{1}{2} e^{-Q_3} \sec^2 Q_2\,(3 \beta -\beta \cos (2 Q_2)+4 \alpha \sin Q_2) \tilde H,
\ee
where ${\cal T}_{f_1}$ is a conformal symmetry in the 2 dimensional space, listed in (\ref{e1h1f1-conf-e1}).
\end{subequations}
These integrals, together with the cubic $J_3=\{J_1,J_2\}$, satisfy the relations
\bea
&&  \{J_1,J_3\} = 4 J_1J_2,\quad  \{J_2,J_3\} = -2 J_2^2+8 P_1^2 \left(2J_1-\alpha \tilde H\right),  \nn\\
&&  J_3^2 = 4 J_1 J_2^2+4 \tilde H\left(4\alpha J_1-\beta^2 \tilde H\right)-16 P_1^2J_1^2.   \nn
\eea

\medskip
When we reduce equation (\ref{e1h1f1eq2}), using $h_1$, the Hamiltonian (\ref{e1h1f1-Ht2h1}) and Casimir (\ref{e1h1f1-cash1}) take the form
\begin{subequations}
\bea
\tilde H &=& \frac{\cos^2(Q_2)}{\alpha+\beta \sin Q_2} \, \left(P_1^2+P_2^2+\frac{1}{4}\, \mbox{sech}^2\, Q_1 P_3^2\right),  \label{e1h1f1eq2DK2}  \\
 J_1 &=&  P_1^2+\frac{1}{4}\, \mbox{sech}^2\, Q_1 P_3^2.  \label{e1h1f1eq2DK2J1}
\eea
This is the $D_4$ kinetic energy, with potential (equivalent to the ``$b_3$'' part of Case B in \cite{03-11}).

It is easily seen that $F_4= h_2 h_4-4 q_1 \sigma H$ (where $\sigma$ is defined in (\ref{e1h1f1eq2-F1})) is a first integral, satisfying $\{H,F_4\}=\{h_1,F_4\}=0$ (see Table 8 of \cite{f19-2} for the full Poisson algebra).  This gives another (independent) quadratic integral for the 2 dimensional system:
\be\label{e1h1f1eq2DK2J2}
J_2 = 4 P_2 \left({\cal T}_{e_1}-{\cal T}_{f_1}\right) + 2  \sec^2 Q_2\, \left(\beta \cos 2 Q_2-3 \beta -4 \alpha \sin Q_2\right) \sinh Q_1 \, \tilde H ,
\ee
\end{subequations}
where ${\cal T}_{e_1}$ and ${\cal T}_{f_1}$ are given in (\ref{e1h1f1-conf-h1}).  These integrals, together with the cubic $J_3=\{J_1,J_2\}$, satisfy
\bea
&&  \{J_1,J_3\} = 4 J_1 J_2,\quad \{J_2,J_3\} = 2 \left(-J_2^2-32 J_1 (6 J_1-P_3^2)+ 16\alpha (8J_1-P_3^2)\tilde H-16 \beta^2 \tilde H^2\right) , \nn\\
&& J_3^2 = 4 \left( J_1 \left(J_2^2+16 J_1 (4 J_1-P_3^2)\right) -16 J_1\tilde H(\alpha (4 J_1+P_3^2)-\beta^2\tilde H)\right)+16 P_3^2\tilde H(8\alpha J_1-\beta^2 \tilde H).\nn
\eea

\subsection{Systems with Isometry Algebra $\left<e_1,e_2,h_2\right>$}\label{2D-super-reduce-e1e2h2}

We give a pair of {\em universal} reductions of the general Hamiltonian in this class:
\be\label{3d-e1e2h2}
H =  \psi(q_3) \left(p_1^2+p_2^2+p_3^2\right).
\ee
We then give the reductions of the two cases presented in Section \ref{3D-super-e1e2h2}.

We give transformations corresponding to $e_1\rightarrow P_1$ and $h_2\rightarrow P_3$, respectively.  Under the involution $\iota_{12}$ we have $e_1\leftrightarrow e_2,\; h_2\rightarrow -h_2$. The first transformation is very simple, just relabelling coordinates so that the conformal factor of the reduced metric is a function of $Q_2$.

\subsubsection{Reduction using the Isometry $e_1\mapsto P_1$}

This transformation is very simple, with
\begin{subequations}
\be \label{e1e2h2-Hte1}
Q_1=q_1,\;\;\; Q_2=q_3,\;\;\; Q_3=q_2 \quad\Rightarrow\quad  \tilde H = \psi(Q_2) \left(P_2^2+P_3^2+P_1^2\right),
\ee
so $P_3$ is a first integral.

Again, we build the 6 dimensional conformal algebra for this 2D metric (kinetic energy).  First note that $e_2,\, e_3,\, h_4,\, e_1f_1+\frac{1}{4} h_1^2,\, e_2 f_2-\frac{1}{2} h_2^2,\,  e_3 f_3-\frac{1}{2} h_3^2$ {\em commute} with $e_1$.
Writing these in terms of $Q_i,P_i$ and discarding the $P_1^2$ components in the quadratic ones, we can derive the following 6 conformal elements:
\bea
&&  {\cal T}_{e_1} = P_3,\;\;\; {\cal T}_{h_1} = -2(Q_2P_2+Q_3P_3),\;\;\; {\cal T}_{f_1}= (Q_2^2-Q_3^2) P_3-2 Q_2Q_3 P_2,  \nn  \\[-1mm]
&&    \label{e1e2h2-conf-e1}    \\[-1mm]
&&  {\cal T}_{e_2} = P_2,\quad {\cal T}_{h_2} = 2 (Q_3P_2-Q_2 P_3),\quad {\cal T}_{f_2}= 2(Q_3^2-Q_2^2) P_2-4 Q_2Q_3 P_3, \nn
\eea
\end{subequations}
which satisfy the relations of $\mathfrak{g}_1 + \mathfrak{g}_2$ in Table \ref{Tab:g1234}, together with the algebraic constraints (\ref{e1h4-conf-constraints}).

\subsubsection{Reduction using the Isometry $h_2\mapsto P_3$}

In view of $\left\{-\frac12\arctan\frac{q_1}{q_2},h_2\right\}=1$ and $\{q_1^2+q_2^2,h_2\}=\{q_3,h_2\}=0$, we consider the generating function
\begin{subequations}
\be\label{e1e2h2-S1h2}
  S= w_1(z,q_3)\;P_1+w_2(q_3)\;P_2+\left(-\frac12\arctan\left(\frac{q_1}{q_2}\right)+w_3(z,q_3)\right)\;P_3,\quad\mbox{where}\quad   z=q_1^2+q_2^2.
\ee
We first require that the coefficients of $P_iP_j$ are zero, giving
$$
  w_2'(q_3)\pa_{q_3}w_1 = w_2'(q_3)\pa_{q_3}w_3  = 0,  \quad 4z\pa_{z}w_1\pa_{z}w_3+\pa_{q_3}w_1\pa_{q_3}w_3=0.
$$
The first two give
$$
w_i=w_i(z)=w_i\left(q_1^2+q_2^2\right),\qquad i=1,3,
$$
since $w_2'(q_3)$ {\em cannot} be zero.  The third then gives
$$
w_1'(z)w_3'(z)=0,\quad\Rightarrow\quad w_3(z)=0,
$$
since $w_1'$ \underline{cannot} be zero, so the canonical transformation (\ref{e1e2h2-S1h2}) now takes the form
\be\label{e1e2h2-S2h2}
   S= w_1(q_1^2+q_2^2)\;P_1+w_2(q_3)\;P_2-\frac12\arctan\left(\frac{q_1}{q_2}\right)\;P_3.
\ee
At this stage, the transformed Hamiltonian is diagonal:
\be\label{e1e2h2-Ht1h2}
 \tilde{H}= 4\psi(q_3) z w_1'^2(z)\;P_1^2+\psi(q_3)w_2'^2(q_3)\;P_2^2+\frac{\psi(q_3)}{4z}\;P_3^2.
\ee
We must choose $w_1$ and $w_2$ so that this Hamiltonian is ``conformal'' in the $P_1-P_2$ components, which requires
$$
  4z w_1'^2(z)=1,\quad w_2'^2(q_3)=1 \quad\Rightarrow\quad w_1=\sqrt{z}=\sqrt{q_1^2+q_2^2},\quad w_2=q_3.
$$
This gives us the final form of the canonical transformation
\be\label{e1e2h2-S3h2}
  S=\sqrt{q_1^2+q_2^2}\;P_1+q_3\;P_2-\frac12\arctan\left(\frac{q_1}{q_2}\right)\;P_3,
\ee
and the Hamiltonian
\be\label{e1e2h2-Ht2h2}
  \tilde{H}= \psi(Q_2)\left(P_1^2+P_2^2+\frac1{4Q_1^2}P_3^2\right).
\ee
This corresponds to a conformally flat metric in the $1-2$ space, defined by $P_3=\mbox{const}$, with $\frac1{4Q_1^2}P_3^2$ corresponding to a potential term.  Since the conformal factor is a function of only $Q_2$, the momentum $P_1$ corresponds to a Killing vector (in 2D), but not a first integral of the entire Hamiltonian.  However, the Casimir function of the original isometry algebra \underline{does} reduce to a first integral of $\tilde H$:
\be\label{e1e2h2-cash2}
e_1^2+e_2^2 = P_1^2+\frac1{4Q_1^2}P_3^2.
\ee
Thus, for \underline{arbitrary} $\psi$, the Hamiltonian $\tilde H$ is Liouville integrable and, indeed, separable.  The form of the potential and of the Casimir are \underline{independent} of the form of $\psi$, so are \underline{universal} properties of this reduction.

Again, we build the 6 dimensional conformal algebra for this 2D metric (kinetic energy).  First note that $e_3,\, h_1,\, f_3,\, e_1^2+e_2^2,\, h_3^2+\frac{1}{4} h_4^2,\,  f_2^2+4 f_1^2$ {\em commute} with $h_2$.
Writing these in terms of $Q_i,P_i$ and discarding the $P_3^2$ components in the quadratic ones, we can derive the following 6 conformal elements:
\bea
&&  {\cal T}_{e_1} = P_1,\;\;\; {\cal T}_{h_1} = -2(Q_1P_1+Q_2P_2),\;\;\; {\cal T}_{f_1}= (Q_2^2-Q_1^2) P_1-2 Q_1Q_2 P_2,  \nn  \\[-1mm]
&&    \label{e1e2h2-conf-h2}    \\[-1mm]
&&  {\cal T}_{e_2} = P_2,\quad {\cal T}_{h_2} = 2 (Q_1P_2-Q_2 P_1),\quad {\cal T}_{f_2}= 2(Q_1^2-Q_2^2) P_2-4 Q_1Q_2 P_1, \nn
\eea
\end{subequations}
which satisfy the relations of $\mathfrak{g}_1 + \mathfrak{g}_2$ in Table \ref{Tab:g1234}, together with the algebraic constraints (\ref{e1h4-conf-constraints}).

\subsubsection{Specific Cases Listed in Section \ref{3D-super-e1e2h2}}

In Section \ref{3D-super-e1e2h2} we gave two specific choices of the function $\psi$, which allowed the addition of quadratic integrals.  To obtain a superintegrable system in the 2 dimensional context, we need two first integrals which commute with the reducing isometry.  One of these can be the Casimir of the isometry algebra, the other coming from the algebra of functions $F_i$.

\subsubsection*{The Case $\psi(z)=\frac{1}{\alpha z  +\beta}$}

When we reduce equation (\ref{e1e2h2eq1}), using $e_1$, the Hamiltonian (\ref{e1e2h2-Hte1}) takes the form
\begin{subequations}
\be\label{e1e2h2eq1DK1}
\tilde H = \frac{1}{\alpha Q_2+\beta} \, \left(P_2^2+P_3^2+ P_1^2\right),
\ee
and the commuting isometry $e_2=P_3$ is a first integral.  This is the $D_1$ kinetic energy, with potential (equivalent to the ``$b_2$'' part of Case 1 in \cite{02-6}).

It can also be seen in Table \ref{Tab:e1e2h2}, that $F_4$ commutes with $e_1$, so can be reduced to
\be\label{e1e2h2eq1DK1J1}
J_1 = P_2 P_3-\frac{1}{2}\,\alpha  Q_3\, \tilde H,
\ee
\end{subequations}
satisfying the relation $\{P_3,J_1\}=\frac{\alpha}{2}\, \tilde H$.

\medskip
When we reduce equation (\ref{e1e2h2eq1}), using $h_2$, the Hamiltonian (\ref{e1e2h2-Ht2h2}) and Casimir (\ref{e1e2h2-cash2}) take the form
\begin{subequations}
\be \label{e1e2h2eq1DK2}
\tilde H = \frac{1}{\alpha Q_2+\beta} \, \left(P_1^2+P_2^2+ \frac{P_3^2}{4 Q_1^2}\right),  \quad    J_1 =  P_1^2+\frac{P_3^2}{4 Q_1^2}.
\ee
This is the $D_1$ kinetic energy, with potential (equivalent to the ``$b_3$'' part of Case 1 in \cite{02-6}).

It is easily seen in Table \ref{Tab:e1e2h2}, that the first integral $2F_1-F_5$ commutes with $h_2$.  This gives another (independent) quadratic integral for the 2 dimensional system:
\be\label{e1e2h2eq1DK2J2}
J_2 = -2 P_2 {\cal T}_{h_1}-\left(4 (\beta+\alpha Q_2) Q_2+\alpha Q_1^2\right)\, \tilde H ,
\ee
where ${\cal T}_{h_1}$ is a conformal symmetry in the 2 dimensional space, listed in (\ref{e1e2h2-conf-h2}).
\end{subequations}
These integrals, together with the cubic $J_3=\{J_1,J_2\}$, satisfy the relations
\bea
&&  \{J_1,J_3\} = -8 \alpha J_1 \tilde H,\quad  \{J_2,J_3\} = 8 \left(\alpha J_2 \tilde H-8 \beta J_1 \tilde H+12 J_1^2\right),  \nn\\
&&  J_3^2 = 16 J_1 \left(4 \beta J_1 \tilde H-\alpha J_2 \tilde H-4 J_1^2\right) -4 \alpha^2\tilde H^2 P_3^2.   \nn
\eea

\subsubsection*{The Case $\psi(z)=\frac{z^2}{\alpha z^2+\beta}$}

When we reduce equation (\ref{e1e2h2eq2}), using $e_1$, the Hamiltonian (\ref{e1e2h2-Hte1}) takes the form
\begin{subequations}
\be\label{e1e2h2eq2DK1}
\tilde H = \frac{Q_2^2}{\alpha Q_2^2+\beta} \, \left(P_2^2+P_3^2+ P_1^2\right),
\ee
and the commuting isometry $e_2=P_3$ is a first integral.  This is the $D_2$ kinetic energy, with potential of ``Type D'' in \cite{03-11}.

It can also be seen that $F_4=e_3h_4+\frac{4\beta q_2}{q_3^2}\, H$ is a first integral, which satisfies $\{e_1,F_4\}=0$, so can be reduced, giving
\be\label{e1e2h2eq2DK1J1}
J_1 = -2P_2 {\cal T}_{h_2} +\frac{4 \beta Q_3}{Q_2^2}\, \tilde H,
\ee
\end{subequations}
where ${\cal T}_{h_2}$ is given in (\ref{e1e2h2-conf-e1}), satisfying the relation $\{P_3,J_1\}= 4\left(\alpha\, \tilde H-P_3^2-P_1^2\right)$.

\medskip
When we reduce equation (\ref{e1e2h2eq2}), using $h_2$, the Hamiltonian (\ref{e1e2h2-Ht2h2}) and Casimir (\ref{e1e2h2-cash2}) take the form
\begin{subequations}
\be \label{e1e2h2eq2DK2}
\tilde H = \frac{Q_2^2}{\alpha Q_2^2+\beta} \, \left(P_1^2+P_2^2+ \frac{P_3^2}{4 Q_1^2}\right),  \quad    J_1 =  P_1^2+\frac{P_3^2}{4 Q_1^2}.
\ee
This is the $D_2$ kinetic energy, with potential (equivalent to the ``$b_3$'' part of ``Type B'' in \cite{03-11}).

It can also be seen that the first integral $F_1$ (see (\ref{e1e2h2eq2-F1})) commutes with $h_2$, giving another (independent) quadratic integral for the 2 dimensional system:
\be\label{e1e2h2eq2DK2J2}
J_2 = P_2 {\cal T}_{f_2}+2\left(\alpha Q_2^2-\frac{\beta Q_1^2}{Q_2^2}\right)\, \tilde H ,
\ee
where ${\cal T}_{f_2}$ is a conformal symmetry in the 2 dimensional space, listed in (\ref{e1e2h2-conf-h2}).
\end{subequations}
These integrals, together with the cubic $J_3=\{J_1,J_2\}$, satisfy the relations
\bea
&&  \{J_1,J_3\} = 16 J_1 \left(\alpha \tilde H - J_1\right),\quad  \{J_2,J_3\} = 16 \left(2 J_1 J_2 -\alpha\tilde H\left(2 \beta \tilde H+J_2\right)+P_3^2\left(J_1-\alpha \tilde H\right)\right),  \nn\\
&&  J_3^2 =  32 J_1 \left(\alpha \tilde H \left(2 \beta \tilde H+J_2\right) - J_1 J_2\right) -16 P_3^2 \left(J_1-\alpha \tilde H\right)^2.   \nn
\eea

\subsection{Systems with Isometry Algebra $\left<h_2,h_3,h_4\right>$}\label{2D-super-reduce-h2h3h4}

We give a {\em universal} reduction of the general Hamiltonian in this class:
\be\label{3d-h2h3h4}
H = \psi\left(q_1^2+q_2^2+q_3^2\right)\, \left(p_1^2+p_2^2+p_3^2\right).
\ee
We then give the reduction of the two cases presented in Section \ref{3D-super-h2h3h4}.

We give one transformation, corresponding to $h_4\rightarrow P_3$.  The transformations using either $h_2$ or $h_3$ are equivalent under the involutions $\iota_{13}$ and $\iota_{12}$, respectively.

We choose $Q_2$ to be a function of $z_2=q_1^2+q_2^2+q_3^2$, which is the common invariant of $h_2, h_3$ and $h_4$ and, as a consequence, the variable which appears in the arbitrary function $\psi$, of the Hamiltonian.

\subsubsection{Reduction using the Isometry $h_4\mapsto P_3$}

In view of $\left\{\frac14\arctan\left(\frac{q_2}{q_3}\right),h_4\right\}=1$ and $\{q_2^2+q_3^2,h_4\}=\{q_1,h_4\}=0$, we consider the generating function
\begin{subequations}
\be\label{h2h3h4-S1}
  S=w_1(z_1,q_1)\;P_1+w_2(z_2)\;P_2+\left(\frac14\arctan\left(\frac{q_2}{q_3}\right)+w_3(z_1,q_1)\right)\;P_3,
\ee
where $z_1=q_2^2+q_3^2$ and $z_2=q_1^2+q_2^2+q_3^2$.

\smallskip
We first require that the coefficients of $P_iP_j$ are zero, giving
$$
w_2'(z_2)\left(2z_1\pa_{z_1} w_1+q_1\pa_{q_1}w_1\right) = w_2'(z_2)\left(2z_1\pa_{z_1} w_3+q_1\pa_{q_1}w_3\right) = 4z_1\pa_{z_1} w_1\pa_{z_1} w_3+\pa_{q_1}w_1\pa_{q_1}w_3 = 0.
$$
The first two give $w_i=w_i\left(\frac{z_1}{q_1^2}\right),\; i=1,3$, since $w_2'(z_2)$ {\em cannot} be zero.  The third then gives $w_1'w_3'=0$.
Since $w_1'$ \underline{cannot} be zero, we have $w_3=0$, so the canonical transformation (\ref{h2h3h4-S1}) now takes the form
\be\label{h2h3h4-S2}
  S= w_1(z_3)\;P_1+s_2(z_2)\;P_2+\frac14\arctan\left(\frac{q_2}{q_3}\right)\;P_3, \quad\mbox{where}\;\;\;  z_3=\frac{q_2^2+q_3^2}{q_1^2}.
\ee
At this stage, the transformed Hamiltonian is diagonal:
\be\label{h2h3h4-Ht1}
  \tilde{H}=4z_3(1+z_3)^2 w_1'^2(z_3)\frac{\psi(z_2)}{z_2}\;P_1^2+4z_2 \psi(z_2)w_2'^2(z_2)\;P_2^2 +\frac{\psi(z_2)}{16z_1}P_3^2.
\ee
We must choose $w_1$ and $w_2$ so that this Hamiltonian is ``conformal'' in the $P_1-P_2$ components, which requires
$$
   4z_3(1+z_3)^2w_1'^2(z_3)=1,\qquad 4z_2w_2'^2(z_2)=\frac1{z_2},
$$
giving
$$
  w_1=\arctan\left(\sqrt{z_3}\right)=\arctan\left(\frac{\sqrt{q_2^2+q_3^2}}{q_1}\right),\quad w_2=\frac12\log{z_2}=\frac12\log{(q_1^2+q_2^2+q_3^2)}.
$$
This gives us the final form of the canonical transformation
\be\label{h2h3h4-S3}
  S= \arctan\left(\frac{\sqrt{q_2^2+q_3^2}}{q_1}\right)\;P_1+\frac12\log{(q_1^2+q_2^2+q_3^2)}\;P_2+\frac14\arctan\left(\frac{q_2}{q_3}\right)\;P_3,
\ee
and the Hamiltonian
\be\label{h2h3h4-Ht2}
  \tilde{H}=\text{e}^{-2Q_2}\psi\left(\text{e}^{2Q_2}\right)\left(P_1^2+P_2^2+\frac{1}{16\sin^2{Q_1}}P_3^2\right).
\ee
This corresponds to a conformally flat metric in the $1-2$ space, defined by $P_3=\mbox{const}$, with $\frac{1}{16} \mbox{cosec}^2{Q_1}\, P_3^2$ corresponding to a potential term.  Since the conformal factor is a function of only $Q_2$, the momentum $P_1$ corresponds to a Killing vector (in 2D), but not a first integral of the entire Hamiltonian.  However, the Casimir function of the original isometry algebra \underline{does} reduce to a first integral of $\tilde H$:
\be\label{h2h3h4-cas}
{\cal C} = \frac{1}{16} (4(h_2^2+h_3^2)+h_4^2) = P_1^2 +\frac{1}{16\sin^2{Q_1}}P_3^2.
\ee
Thus, for \underline{arbitrary} $\psi$, the Hamiltonian $\tilde H$ is Liouville integrable and, indeed, separable.  The form of the potential and of the Casimir are \underline{independent} of the form of $\psi$, so are \underline{universal} properties of this reduction.

Again, we build the 6 dimensional conformal algebra for this 2D metric (kinetic energy).  First note that $e_1,\, h_1,\, f_1,\, e_2^2+e_3^2,\, h_2^2+h_3^2,\,  f_2^2+f_3^2$ {\em commute} with $h_4$.
Writing these in terms of $Q_i,P_i$ and discarding the $P_3^2$ components in the quadratic ones, we can derive the following 6 conformal elements:
\bea
&&  {\cal T}_{e_1} = e^{Q_2} (P_1 \sin Q_1+P_2 \cos Q_1),\quad {\cal T}_{h_1} = 2 P_2,\quad {\cal T}_{f_1}= e^{-Q_2} (P_1 \sin Q_1-P_2 \cos Q_1),  \nn  \\[-1mm]
&&                             \label{h2h3h4-conf}   \\[-1mm]
&&  {\cal T}_{e_2} = e^{Q_2} (P_1 \cos Q_1-P_2 \sin Q_1),\quad {\cal T}_{h_2} = -2 P_1,\quad {\cal T}_{f_2}= 2e^{-Q_2} (P_1 \cos Q_1+P_2 \sin Q_1), \nn
\eea
\end{subequations}
which satisfy the relations of $\mathfrak{g}_1 + \mathfrak{g}_2$ in Table \ref{Tab:g1234}, together with the algebraic constraints (\ref{e1h4-conf-constraints}).

\subsubsection{Specific Cases Listed in Section \ref{3D-super-h2h3h4}}

In Section \ref{3D-super-h2h3h4} we gave two specific choices of the function $\psi$, which allowed the addition of quadratic integrals.  To obtain a superintegrable system in the 2 dimensional context, we need two first integrals which commute with the reducing isometry.  One of these can be the Casimir of the isometry algebra, the other coming from the algebra of functions $F_i$.

\subsubsection*{The Case $\psi(z)=\frac{\sqrt{z}}{\alpha \sqrt{z}  +\beta}$}

When we reduce equation (\ref{h2h3h4eq1}), using $h_4$, the Hamiltonian (\ref{h2h3h4-Ht2}) takes the form
\begin{subequations}
\be\label{h2h3h4eq1DK1}
\tilde H = \frac{e^{-Q_2}}{\alpha e^{Q_2}+\beta} \, \left(P_1^2+P_2^2+\frac{P_3^2}{16 \sin^2 Q_1}\right),
\ee
which is the $D_3$ kinetic energy, with a potential.

We already have the integral $J_1$, derived from the Casimir of the isometry algebra (see (\ref{h2h3h4-cas})).
It can also be seen in Table \ref{Tab:h2h3h4:1}, that $F_1$ commutes with $h_4$, so can be reduced to
\be\label{h2h3h4eq1DK1J2}
J_2 = 2 P_2 {\cal T}_{f_1}+\left(\beta+2 \alpha e^{Q_2}\right)\,\cos Q_1 \, \tilde H,
\ee
\end{subequations}
where ${\cal T}_{f_1}$ is a conformal symmetry in the 2 dimensional space, listed in (\ref{h2h3h4-conf}).

These integrals, together with the cubic $J_3=\{J_1,J_2\}$, satisfy the relations
\bea
&&  \{J_1,J_3\} = -4 J_1 J_2,\quad  \{J_2,J_3\} = 2 J_2^2-\tilde H \left(2 \beta^2 \tilde H+16 \alpha J_1-\frac{a}{2} P_3^2\right),  \nn\\
&&  J_3^2 = -4 J_1 J_2^2 +\frac{1}{4} \tilde H \left(16 J_1-P_3^2\right) \left(\beta^2 \tilde H+4 \alpha J_1\right).   \nn
\eea

\subsubsection*{The Case $\psi(z)=\frac{1}{\alpha z  +\beta}$}

When we reduce equation (\ref{h2h3h4eq2}), using $h_4$, the Hamiltonian (\ref{h2h3h4-Ht2}) takes the form
\begin{subequations}
\be\label{h2h3h4eq2DK1}
\tilde H = \frac{e^{-2Q_2}}{\alpha e^{2Q_2}+\beta} \, \left(P_1^2+P_2^2+\frac{P_3^2}{16 \sin^2 Q_1}\right),
\ee
which is again the $D_3$ kinetic energy, with potential (equivalent to the ``$b_2$'' part of Case B in \cite{03-11}).

We already have the integral $J_1$, derived from the Casimir of the isometry algebra (see (\ref{h2h3h4-cas})).
It can also be seen that $F_1$ commutes with $h_4$ (see Table 10 of \cite{f19-2}, for the full Poisson algebra), so can be reduced to
\be\label{h2h3h4eq2DK1J2}
J_2 = {\cal T}_{f_1}^2 -\alpha e^{2Q_2}\,\cos^2 Q_1 \, \tilde H,
\ee
\end{subequations}
where ${\cal T}_{f_1}$ is a conformal symmetry in the 2 dimensional space, listed in (\ref{h2h3h4-conf}).

These integrals, together with the cubic $J_3=\{J_1,J_2\}$, satisfy the relations
\bea
&&  \{J_1,J_3\} = -16 J_1 J_2 +\frac{1}{2}\,\beta \tilde H \left(16 J_1-P_3^2\right),\quad  \{J_2,J_3\} = 8 J_2 \left(J_2-\beta\tilde H\right) -\alpha  \tilde H \left(16 J_1-P_3^2\right),  \nn\\
&&  J_3^2 = -16 J_1 J_2^2 +\frac{1}{16} \tilde H \left(16 J_1-P_3^2\right) \left(16 \alpha J_1+16 \beta J_2- \alpha P_3^2\right).   \nn
\eea

\section{Reduction to 2 Dimensional Systems: 4D Isometry Algebras}\label{4D-super-reduce}

We can see in Table \ref{Tab:SubAlg-Ham} that there are two {\em genuinely} 4D isometry algebras which, therefore, do not correspond to flat or constant curvature metrics.  Having 4 isometries {\em completely determines} the Hamiltonian, so there are no arbitrary functions $\psi$.

\subsection{The Algebra $\langle e_1,h_1,f_1\rangle\oplus\langle h_4\rangle$}

It can be seen in Table \ref{Tab:g1234} that $\langle e_1,h_1,f_1\rangle$ is just the algebra $\mathfrak{g}_1$ (a copy of $\mathfrak{sl}(2)$) and that $h_4$ commutes with $\mathfrak{g}_1$.  This means that we can {\em simultaneously straighten} the pairs $(e_1,h_4)$ or $(h_1,h_4)$ (with $(f_1,h_4)$ being related to $(e_1,h_4)$ through $\iota_{ef}$).

The general Hamiltonian with this isometry algebra is given by a Casimir
\be\label{e1h1f1h4-eq}
H=(q_2^2+q_3^2)(p_1^2+p_2^2+p_3^2)=e_1f_1+\frac14h_1^2+\frac1{16}h_4^2.
\ee
The more general Casimir, with $\gamma h_4^2$, has terms $p_i p_j$.

This can be considered as a restriction of the cases of several smaller algebras listed in Table \ref{Tab:SubAlg-Ham}, with the forms of the function $\psi$.  Clearly $\langle e_1,h_4\rangle$ is a subalgebra and this restriction is achieved through $\psi(z)=z$.  Similarly $\langle e_1,h_1,f_1\rangle$ is a subalgebra and this restriction corresponds to $\psi(z)=1+z^2$.  The subalgebra $\left<h_1,h_2\right>$ is equivalent to $\left<h_1,h_4\right>$ under the involution $\iota_{13}$ and this restriction corresponds to $\psi(z)=1$.

\subsubsection{Reductions with the Pair $(e_1,h_4)$}

Here we can use the canonical transformations of Section \ref{2D-super-reduce-e1h4} to set $(e_1,h_4)\mapsto (P_1,P_3)$.

The canonical transformation (\ref{TRe1h4:1}) reduces (\ref{e1h1f1h4-eq}) to
\be\label{e1h1f1h4-eq-e1red}
  \tilde{H}=\frac1{16}\left(P_2^2+P_3^2+16 \text{e}^{8Q_2}P_1^2\right),
\ee
corresponding to a {\em flat} metric.  This is a consequence of two commuting Killing vectors in the 2 dimensional domain.  Since $e_1$ commutes with $h_4$ {\em and} the $\mathfrak{sl}(2)$ Casimir $e_1f_1+\frac14h_1^2$, these can be reduced:
$$
\left(h_4,e_1f_1+\frac14h_1^2\right) \mapsto \left(P_3,\frac1{16}\left(P_2^2+16 \text{e}^{8Q_2}P_1^2\right)\right),
$$
giving the Noether constants $(P_2,P_3)$.

The canonical transformation (\ref{TRe1h4:2}) reduces (\ref{e1h1f1h4-eq}) to
\be\label{e1h1f1h4-eq-h4red}
  \tilde{H}=Q_2^2\left(P_1^2+P_2^2+\frac{1}{16} P_3^2\right),
\ee
corresponding to a {\em constant curvature} metric.  This is a consequence of the existence of three Killing vectors in the 2 dimensional domain, with $\mathfrak{sl}(2)$ relations.  Since $h_4$ commutes with $\mathfrak{g}_1$, each element can be reduced:
$$
(e_1,h_1,f_1) \mapsto \left(P_1,-2(Q_1P_1+Q_2P_2),(Q_2^2-Q_1^2)P_1^2-2 Q_1Q_2P_2\right),
$$
which satisfy the same Poisson relations as $\mathfrak{g}_1$.

\subsubsection{Reductions with the Pair $(h_1,h_4)$}

Here we can use the canonical transformations of Section \ref{2D-super-reduce-h1h2} to set $(h_1,h_4)\mapsto (P_1,P_3)$.

The canonical transformation (\ref{TRh1h2:h1}), under $\iota_{13}$, gives us
\begin{subequations}
\be\label{TRh1h4:h1}
S = -\frac{1}{4}\,\log\left(q_1^2+q_2^2+q_3^2\right)\,P_1+\frac14\, \log\left(\frac{q_1+\sqrt{q_1^2+q_2^2+q_3^2}}{\sqrt{q_2^2+q_3^2}}\right)\,P_2-\frac14\arctan\left(\frac{q_3}{q_2}\right)\,P_3,
\ee
and reduces (\ref{e1h1f1h4-eq}) to
\be\label{e1h1f1h4-eq-h1red}
  \tilde{H}=\frac1{16}\left(P_2^2+P_3^2+ 4 \mbox{sech}^2 4Q_2 \, P_1^2\right),
\ee
\end{subequations}
corresponding to a {\em flat} metric.  This is a consequence of two commuting Killing vectors in the 2 dimensional domain.  Since $h_1$ commutes with $h_4$ {\em and} the $\mathfrak{sl}(2)$ Casimir $e_1f_1+\frac14h_1^2$, these can be reduced:
$$
\left(h_4,e_1f_1+\frac14h_1^2\right) \mapsto \left(P_3,\frac1{16}(P_2^2+ 4 \mbox{sech}^2 4Q_2 \, P_1^2)\right),
$$
giving the Noether constants $(P_2,P_3)$.

The canonical transformation (\ref{TRh1h2:h2}), under $\iota_{13}$, gives us
\begin{subequations}
\be\label{TRh1h4:h4}
S =-\frac{1}{4}\,\log{\left(q_1^2+q_2^2+q_3^2\right)}\,P_1+\frac12\arctan\left(\frac{q_1}{\sqrt{q_2^2+q_3^2}}\right)\,P_2-\frac14\arctan\left(\frac{q_3}{q_2}\right)\,P_3,
\ee
and reduces (\ref{e1h1f1h4-eq}) to
\be\label{e1h1f1h4-eq-h4red2}
  \tilde{H}= \frac{1}{4} \cos^2 2 Q_2\, \left(P_1^2+P_2^2+\frac{1}{4} \sec^2 2Q_2 \, P_3^2\right),
\ee
\end{subequations}
corresponding to a {\em constant curvature} metric.  This is a consequence of the existence of three Killing vectors in the 2 dimensional domain, with $\mathfrak{sl}(2)$ relations.  Since $h_4$ commutes with $\mathfrak{g}_1$, each element can be reduced:
$$
(e_1,h_1,f_1) \mapsto \left(\frac{1}{2}\, e^{2 Q_1} (P_2 \cos 2Q_2 -P_1 \sin 2 Q_2),P_1,\frac{1}{2}\, e^{-2 Q_1} (P_2 \cos 2Q_2 +P_1 \sin 2 Q_2)\right),
$$
which satisfy the same Poisson relations as $\mathfrak{g}_1$.

\br[Direct Transformation]
The system (\ref{e1h1f1h4-eq-h4red}) can be directly related to (\ref{e1h1f1h4-eq-h4red2}) within the 2 dimensional domain (by composing the above canonical transformations):
$$
\bar Q_1 = e^{-2 Q_1} \sin 2Q_2,\quad \bar Q_2 = e^{-2 Q_1} \cos 2Q_2,\quad \bar Q_3 = Q_3.
$$
\er

\subsection{The Algebra $\langle h_2,h_3,h_4\rangle\oplus\langle h_1\rangle$}

It can be seen in Table \ref{Tab:g1234} that $\langle h_2,h_3,h_4\rangle$ is just the algebra $\mathfrak{so}(3)$ and that $h_1$ commutes with this algebra.  This means that we can {\em simultaneously straighten} the pair $(h_1,h_2)$ (with the pairs $(h_1,h_3)$ and $(h_1,h_4)$ being related through $\iota_{23}$ and $\iota_{13}$ respectively).

The general Hamiltonian with this isometry algebra is given by a Casimir
\be\label{h2h3h4h1-eq}
H=(q_1^2+q_2^2+q_3^2)(p_1^2+p_2^2+p_3^2)= \frac14 h_1^2+\frac14 \left(h_2^2+h_3^2+\frac14 h_4^2\right).
\ee
The more general Casimir, with $\gamma h_1^2$, has terms $p_i p_j$.

This can also be considered as a restriction of the cases of some smaller algebras listed in Table \ref{Tab:SubAlg-Ham}, with the forms of the function $\psi$.  Clearly $\langle h_1,h_2\rangle$ is a subalgebra and this restriction is achieved through $\psi(z)=1+z$.  Similarly $\langle h_2,h_3,h_4\rangle$ is a subalgebra and this restriction corresponds to $\psi(z)=z$.

\subsubsection{Reductions with the Pair $(h_1,h_2)$}

Here we can use the canonical transformations of Section \ref{2D-super-reduce-h1h2} to set $(h_1,h_2)\mapsto (P_1,P_3)$.

The canonical transformation (\ref{TRh1h2:h1}) reduces (\ref{h2h3h4h1-eq}) to
\be\label{h2h3h4h1-eq-h1red}
  \tilde{H}=\frac14 \cosh^2 2 Q_2\left(P_2^2+P_3^2+ \mbox{sech}^2 2 Q_2\, P_1^2\right),
\ee
corresponding to a {\em constant curvature} metric.  This is a consequence of the existence of three Killing vectors in the 2 dimensional domain, with $\mathfrak{so}(3)$ relations.  Since $h_1$ commutes with $\mathfrak{so}(3)$, each element can be reduced:
$$
(h_2,h_3,h_4) \mapsto \left(P_3, P_2 \cosh 2Q_2 \cos 2 Q_3 +P_3 \sinh 2 Q_2 \sin 2 Q_3,2 P_3 \sinh 2 Q_2 \cos 2 Q_3-2 P_2 \cosh 2Q_2 \sin 2 Q_3 \right),
$$
which satisfy the same Poisson relations as $(h_2,h_3,h_4)$ in Table \ref{g1234}.

The canonical transformation (\ref{TRh1h2:h2}) reduces (\ref{h2h3h4h1-eq}) to
\be\label{h2h3h4h1-eq-h2red}
  \tilde{H}= \frac14 \left(P_1^2+P_2^2+\frac{1}{4} \sec^2 2 Q_2\, P_3^2\right),
\ee
corresponding to a {\em flat} metric.  This is a consequence of two commuting Killing vectors in the 2 dimensional domain.  Since $h_2$ commutes with $h_1$ {\em and} the $\mathfrak{so}(3)$ Casimir $h_2^2+h_3^2+\frac14 h_4^2$, these can be reduced:
$$
\left(h_1,h_2^2+h_3^2+\frac14 h_4^2\right) \mapsto \left(P_1,P_2^2+\frac{1}{4} \sec^2 2 Q_2\, P_3^2\right),
$$
giving the Noether constants $(P_1,P_2)$.

\section{Conclusions}

This paper has continued our work of \cite{f18-1,f19-3,f19-2} on building higher order integrals out of conformal symmetries, and, more generally, building Poisson algebras related to superintegrable systems.  We emphasise that whilst {\em all} higher order integrals can be built out of Killing vectors in the constant curvature case, there {\em was} no such algebraic method available for more general spaces.  The method used here, described in Section \ref{quadrat-int}, is the only one available for general conformally flat spaces.

In this paper we have constructed a class of superintegrable Hamiltonians, representing geodesic flows on a conformally flat manifold.  We specifically constructed systems with 2, 3 or 4 Killing vectors, but no more (wishing to avoid constant curvature spaces) and then used the conformal algebra to build further quadratic integrals, corresponding to rank 2 Killing tensors, and derived the full Poisson algebra of first integrals.  By adapting coordinates to different Killing vectors we derived {\em universal reductions} for Hamiltonians, corresponding to each specific isometry algebra.  In particular, superintegrable geodesic systems in 3 degrees of freedom were reduced to Hamiltonians with Darboux-Koenigs kinetic energy and a potential function associated with the original isometry algebra.

In this paper we have not considered the addition of potential functions, which can lead to very complicated Poisson algebras in the case of 3 degrees of freedom (see \cite{f18-1}).  However, imposing invariance under one or more of the isometries of the kinetic energy can simplify the calculations.  For example, a fully rotationally invariant potential can be added to the Hamiltonian (\ref{h2h3h4eq1}) and this gives a curved space generalisation of the Kepler problem, with the extended $F_i$ just being generalisations of the Runge-Lenz integrals.  This topic is left for a future paper.

An important, but difficult, problem is to construct higher (than second) order integrals.  Some third order integrals were considered in \cite{f19-3}, where one of the cases of \cite{11-3} was constructed, but even in 2 degrees of freedom this was complicated.  This general problem is discussed in \cite{17-4}, for systems in 2 degrees of freedom, with one Killing vector.

Another important extension is the study of systems with more degrees of freedom.  This is generally a difficult task, but our approach should be computationally simpler.  Certainly, there are specific classes of Hamiltonian which could be studied.  In particular, this approach has been used to construct a new invariant in the context of the Eisenhart lift in General Relativity \cite{f19-1}.

\subsection*{Acknowledgements}

This work was supported by NSFC Grant No. 11871396 and NSF of Shaanxi Province of China, Grant No. 2018JM1005.


\end{document}